\documentclass[fleqn,usenatbib]{mnras}

\usepackage{newtxtext,newtxmath}
\usepackage{lscape}
\usepackage{comment}

\usepackage[T1]{fontenc}
\usepackage{ae,aecompl}

\DeclareRobustCommand{\VAN}[3]{#2}
\let\VANthebibliography\thebibliography
\def\thebibliography{\DeclareRobustCommand{\VAN}[3]{##3}\VANthebibliography}

\usepackage{graphicx}	
\usepackage{amsmath}	

\usepackage{xspace}
\usepackage{enumitem}
\usepackage{orcidlink}

\graphicspath{{./figs/}}

\newcommand{\lmxb}{{\it J1858}} 

\newcommand{\src}{{\it J1858}}

\title[The donor and evolution of {\it Swift J1858.6--0814}]
{Shedding far-ultraviolet light on the donor star and evolutionary state of the neutron-star LMXB {\it Swift J1858.6--0814}.}

\author[N. Castro Segura et al.]{N. Castro Segura$^{\orcidlink{0000-0002-5870-0443}}$,$^{1}$\thanks{E-mail: Noel.Castro-Segura@soton.ac.uk}
C~Knigge$^{\orcidlink{0000-0002-1116-2553}}$,$^{1}$
J.~H.~Matthews$^{\orcidlink{0000-0002-3493-7737}}$,$^{2}$
F. M. Vincentelli,$^{3,4}$
P. Charles,$^{1}$
K. S. Long,$^{5,6}$
\newauthor
D. Altamirano,$^{1}$
D. A. H. Buckley$^{\orcidlink{
[0000-0002-7004-9956}}$,$^{7,8,9,10}$
D. Modiano,$^{11}$
M. A. P. Torres,$^{3,4}$
D. J. K. Buisson,$^{12}$
S. Fijma,$^{11}$
\newauthor
K. Alabarta,$^{13}$
N. Degenaar,$^{11}$ 
M. Georganti,$^{1}$
M. C. Baglio,$^{13,14}$
\newauthor
\emph{\normalsize Affiliations are listed at the end of the paper}
}

 \date{Accepted 2023 September 28. Received 2023 September 13; in original form 2023 June 5}

\pubyear{2023}

\begin{document}
\label{firstpage}
\pagerange{\pageref{firstpage}--\pageref{lastpage}}
\maketitle

\begin{abstract}
The evolution of accreting X-ray binary systems is closely coupled to the properties of their donor stars. As a result, we can constrain the evolutionary track a system is by establishing the nature of its donor. Here, we present far-UV spectroscopy of the transient neutron-star low-mass X-ray binary \src\ in three different accretion states (low-hard, high-hard and soft). All of these spectra exhibit anomalous N\,{\sc v}, C\,{\sc iv},  Si\,{\sc iv} and He\,{\sc ii} lines, suggesting that its donor star has undergone CNO processing.  We also determine the donor's effective temperature,  $T_{d} \simeq 5700$~K, and radius, $R_d \simeq 1.7~R_{\odot}$, based on photometric observations obtained during quiescence. Lastly, we leverage the transient nature of the system to set an upper limit of $\dot{M}_{\rm acc} \lesssim 10^{-8.5}~M_{\odot}~yr^{-1}$ on the present-day mass-transfer rate. Combining all these with the orbital period of the system, $P_{\rm orb} = 21.3$~hrs, we search for viable evolution paths. The initial donor masses in the allowed solutions span the range $1~M_{\odot} \lesssim M_{d,i} \lesssim 3.5~M_{\odot}$. All but the lowest masses in this range are consistent with the strong CNO-processing signature in the UV line ratios. The present-day donor mass in the permitted tracks are $0.5~M_{\odot}\lesssim M_{d,obs} \lesssim 1.3~M_{\odot}$, higher than suggested by recent eclipse modelling. 
Since $P_{\rm orb}$ is close to the so-called bifurcation period, both converging and diverging binary tracks are permitted. If \src\ is on a converging track, it will end its life as an ultra-compact system with a sub-stellar donor star.

\end{abstract}

\begin{keywords}
accretion, accretion discs -- stars: neutron -- stars: evolution --  X-rays: binaries -- ultraviolet: stars --  binaries: eclipsing
\end{keywords}



\section{Introduction}

In low-mass X-ray binaries (LMXBs), a neutron star (NS) or black-hole (BH) accretes material via a disc from a Roche-lobe-filling, low-mass companion. During the accretion process, gravitational potential energy is released in the form of radiation across the electromagnetic spectrum, but especially in the X-ray regime \citep{FrankKingRaine2002apa..book.....F}. If the mass-transfer rate from the donor is high enough to maintain the disc in a fully ionized state, the accretion process is stable. Systems in this state are luminous persistent X-ray sources. However, the mass-transfer rates in many LMXBs are lower than this, making their discs susceptible to a viscous-thermal instability \citep{Lasota2016}. These systems are characterized by long periods of quiescence (during which mass accumulates in the disc), punctuated by violent outbursts (during which mass is rapidly transferred onto the compact accretor). Such LMXBs are therefore transient X-ray sources \citep{CharlesCoe2006csxs.book..215C,XRBs_XRTE_era2015ApJ...805...87Y,Alabarta_failed_outbursts2021MNRAS.507.5507A}.

The massive progenitors of NSs and BHs expand to radii as large as $R \simeq 1000~R_{\odot}$ during their evolution. By contrast, the short orbital periods of LMXBs ($P_{\rm orb} \lesssim 10$~d) correspond to present-day binary separations $a \lesssim 25~R_{\odot}$. This implies that the formation of LMXBs must have involved a common-envelope (CE) phase that dramatically reduced the orbital period and binary separation of an initially much wider binary system. A semi-detached LMXB phase can then be triggered and driven in one of two ways. First, the non-degenerate companion can expand to fill its Roche Lobe after it leaves the main sequence. This is referred to as Case A, B or C mass transfer, depending on whether the donor is on the sub-giant, red giant or asymptotic giant branch. Second, angular momentum loss (AML) from the system can shrink the binary orbit to the point where the Roche Lobe makes contact with the stellar radius. The requisite AML can either be due to a magnetically channeled wind from the companion (``magnetic braking'') or, at the shortest orbital periods, gravitational radiation \citep[see e.g.][ for a review]{Belloni_BinaryEvo_review2023}. 

In general, it is difficult to identify the specific evolution track for an LMXB. Systems with very different initial companion masses can reach similar present-day LMXB configurations. This makes it hard to be sure about the make-up and size of the overall LMXB population. From an evolutionary point of view, the most constraining characteristics of an LMXB -- apart from its orbital period -- are the properties of its donor star. This is partly because the donor typically {\em drives} the evolution (via radius expansion or magnetic braking), and partly because its properties can change significantly during the evolution (due to the mass loss it experiences). Since the CNO cycle only operates in stars with masses $M \gtrsim 1.4~M_{\odot}$, the abundance signature of CNO processing is a powerful way to identify systems that have evolved via an intermediate-mass X-ray binary (IMXB) phase \citep[e.g.][]{podsi2002,Haswell2002MNRAS.332..928H,Gansicke2003ApJ...594..443G,Froning2011ApJ...743...26F,Froning2014ApJ...780...48F}. 

 {\it Swift~J1858.6-0814} (Hereafter, \src) was detected as a transient new X-ray source with the {\it Neil Gehrels Swift Observatory} \citep{swift} on October 2018. During its $\simeq 1.5$~yr outburst, it exhibited strong variability across the electromagnetic spectrum, as well as evidence for powerful outflows \citep[e.g. ][]{J1858Vasilopoulos2018ATel12164....1V,1858X-rwinds2020MNRAS.498...68B,1858_optical_winds,1858_radio_Eijnden2020MNRAS.496.4127V,Castro-Segura2022Natur.603...52C}. Its peculiar flaring behaviour is both qualitatively and quantitatively reminiscent of that observed in BH-LMXBs accreting at super-Eddington rates \citep{J1858NusHare2020ApJ...890...57H,MottaV404Cyg_b2017MNRAS.471.1797M,Kimura2016Natur.529...54K,Chaty2003MNRAS.343..169C,Vincentelli2023Natur.615...45V}. However, the discovery of thermonuclear-powered Type-I X-ray bursts established unambiguously that the accretor in \src\ is a NS  \citep{J1858SoftBuisson2020ATel13536....1B}. Since Type-I bursts are Eddington-limited, they can also be used to estimate the source distance, yielding $d = 12.8 \pm 0.7$~kpc \citep{1858X-rayBursts2020MNRAS.499..793B} for \src . Moreover, the system turns out to be deeply eclipsing, providing an extremely precise orbital period estimate ($P_{\rm orb} \simeq 21.3$~hrs), as well as strong constraints on the mass ratio and inclination \citep{1858_Eclipses2021MNRAS.503.5600B}.

During the 2019-2021 outburst of \src, we used the {\em Cosmic Origins Spectrograph} (COS) and {\em Space Telescope Imaging Spectrograph} (STIS) onboard the {\em Hubble Space Telescope} (HST) to obtain time-resolved, far- and near-UV spectra of the system in three distinct spectral states. During the first epoch (program ID 15984), \src\ was in a luminous hard/flaring state. Our initial analysis of this data set revealed the blue-shifted absorption signatures of a "warm" disc wind in the persistent (non-variable) component of several far-UV resonance lines \citep{Castro-Segura2022Natur.603...52C}. These far-UV signatures were observed simultaneously with optical signatures of a cool wind component. The second and third observing epochs took place during the soft and low-hard states, respectively (program ID 16066). 

Here, we present time-averaged far- and near-UV spectra for all three epochs. All of them exhibit anomalous emission line spectra featuring strong N\,{\sc v}\,1240\,\AA, yet barely detectable or absent C\,{\sc iv}\,1550\,\AA. This points to line formation taking place in material that has undergone CNO processing, and hence to an initial donor mass in excess of $\simeq 1.4~M_{\odot}$. By combining this constraint with the spectral energy distribution of the donor -- and the orbital period of the system -- we are able identify the family of binary evolution tracks that might have produced \src\ \citep{DABS_Mangat2023A&C....4200681M}.

\section{Observations}\label{sec: obs}

\subsection{X-ray light curve}
\begin{figure*}
      \centering
      \includegraphics[width=0.98\textwidth]{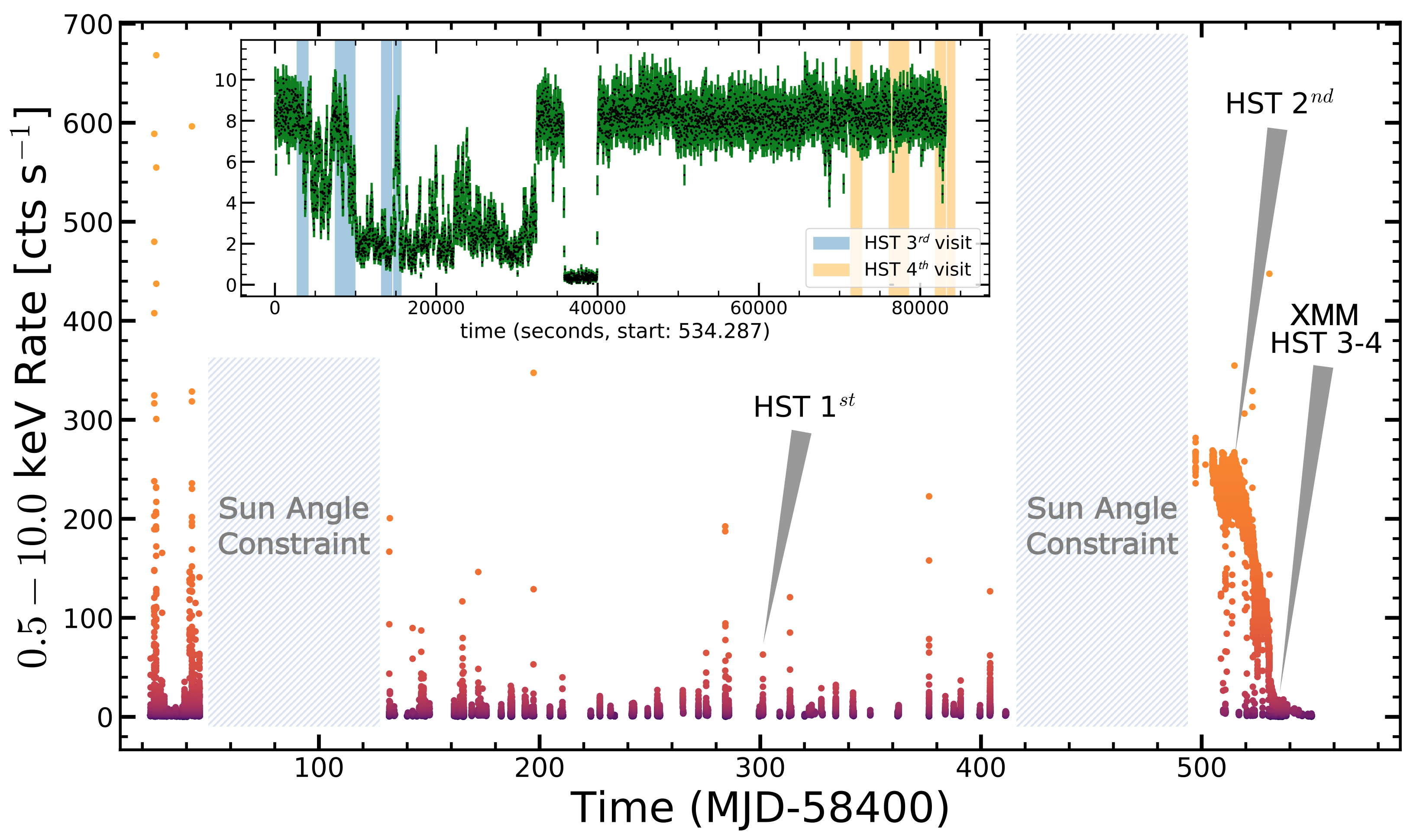}
      \caption[]{X-Ray light curve overview of the outburst as observed with NICER in the 0.5–10keV band \citep[reproduced from ][]{1858X-rayBursts2020MNRAS.499..793B}. The two large gaps are caused by Sun constraints. The source exhibits flares during the first 450 days. Some of these flares reach the eddington limit. This period has been associated with the canonical hard state \citep{1858_Eclipses2021MNRAS.503.5600B}. The time of the HST visits are indicated with the labelled triangles. Colour code refers to the observed count rate. The inset shows the count-rate in the same energy range as the main figure of the simultaneous XMM-Newton observation, it samples a period of absorption dips that happen before an eclipse. This observation covers the last two visits HST visits (HST 3-4), the span during the sub-exposures for each visit is indicated as a coloured shaded region.} 
      \label{fig: full lc}
    \end{figure*}

To provide context to when the spectroscopy were taken, in Figure \ref{fig: full lc} we reproduce the full outburst light curve shown in \citet{1858X-rayBursts2020MNRAS.499..793B} and \citet{Castro-Segura2022Natur.603...52C} gathered with NICER \citep{NICER}. The inset shows the light curve constructed from the {\it XMM-Newton} observations \citep{XMM-Newton}, obtained during the third observing epoch, which took place in the low-hard state.  Two HST visits were carried out during this epoch (HST-3 and HST-4). The times and duration of the far-UV exposures associated with these visits are indicated with shaded regions in the inset of Figure~\ref{fig: full lc}.

We extracted the XMM-Newton lightcurve in the 0.5-10 keV band from the Timing mode observation performed on the 2021 March 26$^{th}$ (OBSID: 0865600201, PI Castro Segura). We  filtered events between column {\sc RAWX} 27 and 47, with {\sc PATTERN$<=4$} and {\sc FLAG$==0$}. Barycentric correction to the event timestamps was applyed to through the SAS software {\sc barycen}, and then we rebinned the lightcurve to 10s time resolution.

\subsection{Ultraviolet spectroscopy}
\label{sec: uv_spec}

We executed a total of four visits in three different spectral states on \src\ with HST. At the time of the $1^{st}$ ($\rm MJD \simeq 58701.1$), the source was in the high-hard/flaring state. This visit included two orbits of far-UV spectroscopy with COS \citep[][]{COS} and one orbit of near-UV spectroscopy with STIS \citep[][]{STIS}, only for the latter orbit, target acquisition was carried out via a blind offset from a nearby star. The $2^{nd}$ visit was performed when the source emerged from the sun glare in a renewed steady soft state \citep{J1858SoftBuisson2020ATel13536....1B}. Three consecutive orbits of
far-UV spectroscopy were obtained during this visit, starting at $\rm MJD \simeq 58914.8$. The last two visits were performed simultaneously with {\it XMM-Newton} (OBS ID 0865600201) when the source transitioned to the low-hard state. Each of these two visits spanned 3 orbits. The visits were separated by $\rm \sim 19~hrs$, with the first observation starting at $\rm MJD \simeq 58934.3$. Once again, only far-UV observations with COS were carried out during these visits. We used the G140L grating with the primary science aperture (PSA) in the LP4 position for all of our COS far-UV observations, yielding a spectral resolving power of $\mathrm{R = \lambda / \Delta \lambda} \sim 900$. 

For the STIS NUV observations in visit 1, we used the G230L grating with the 0.2 arcsec slit, yielding $\mathrm{R  \sim 900}$. We note that the recorded flux level in the G230L spectrum was $\simeq 5\times$ lower than the one in the G140L spectrum obtained in the same visit (in the wavelength region common to both spectra). However, this is probably an artefact associated with the mis-centering of \src in the small STIS/G230L slit following the blind offset target acquisition. In line with this, the cross-dispersion profile of the 2-D NUV spectrum is anomalously asymmetric. It is thus unlikely that the low NUV flux level is due to intrinsic source variability. In any case, we only use the NUV data to estimate the reddening towards the \src and confirm the absence of Mg~{\sc ii}~2800~\AA line. Neither of these depends on the absolute flux level.

 All of the data were reduced using the HST pipelines {\em calstis} and {\em calcos} for the near- and far-UV,  respectively\footnote{Provided by The Space Telescope Science Institute (\url{https://github.com/spacetelescope})}. In order to estimate the line fluxes associated with the most important far-UV transitions (N\,{\sc v}, Si\,{\sc iv}, He\,{\sc ii} and C\,{\sc iv}), 
  we first corrected the spectra for interstellar extinction and fitted a first order polynomial to the continuum adjacent to each line. We then estimated each line flux by numerically integrating over the continuum-subtracted spectrum. 
  
  The associated uncertainties were estimated via Monte Carlo simulations. We first obtained the standard deviations, $\sigma_{\rm obs}$, of the continuum-subtracted flux values in the continuum bands associated with each line. The local continuum around each of the lines was selected by visual inspection to ensure there is no contamination from nearby bound-bound transitions.\footnote{The continuum regions selected are: $\lambda = 1226.6\, $\AA$ - 1231.5\, $\AA\ and $\lambda = 1252.6 6\, $\AA$ - 1255 6\, $\AA\ for N~{\sc v}, $\lambda = 1381\,$\AA$-1387\,$\AA\ and $\lambda= 1415.6\,$\AA$-1426\,$\AA\ for Si~{\sc iv}, $\lambda = 1546.5\,$\AA$-1548\,$\AA\ and $\lambda=1553\,$\AA$-1555.5\,$\AA\ for C~{\sc iv}, $\lambda = 1615\,$\AA$-1627\,$\AA\ and $\lambda= 1648\,$\AA$-1654\,$\AA\ for He~{\sc ii}.}. For each line, we then created mock data sets by replacing each continuum-subtracted flux, $F_{\lambda, {\rm obs}}$, with a random value draw from a Gaussian distribution with mean $F_{\lambda, {\rm obs}}$ and standard deviation $\sigma_{\rm obs}$. The line flux for each mock data set was then obtained by once again integrating numerically across the (mock) line profile. For each line, this procedure was repeated 5000 times, and the error was taken to be the standard deviation of the 5000 mock line flux estimates. The resulting line flux measurements and uncertainties are presented in Table~\ref{table: lines}.

\begin{figure*}
      \centering
      \includegraphics[width=0.98\textwidth]{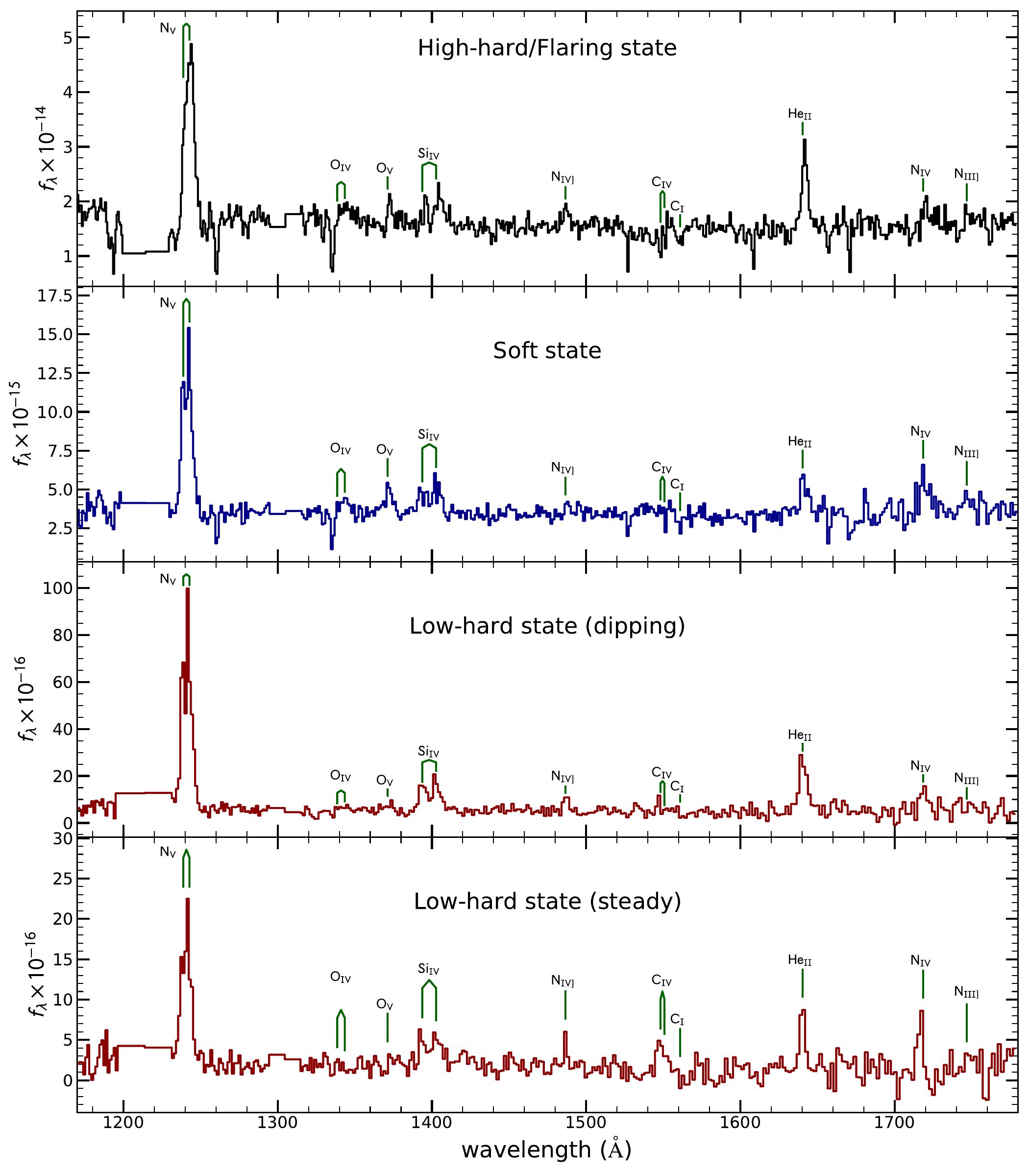}
      \caption[]{ The spectrum of \src\ as observed with HST at three different epochs corrected for extinction. The spectral states from top to bottom are high-hard, soft and low hard state. The positions of C, N, O, He and Si emission lines are indicated with a tick and corresponding label. The extreme N to C ratio suggests that the accreted material has undergone CNO processing. 

      \label{fig: fuv spectra}}
    \end{figure*}

In all visits across all different spectral states, N\,{\sc v} is by far the most prominent emission line in the far-UV spectrum, followed by He\,{\sc ii} (see Fig.~\ref{fig: fuv spectra}). By contrast, C\,{\sc iv} is mainly in absorption, with the exception of the 4$^{th}$ visit during the steady, low-hard state.  Si\,{\sc iv} and O\,{\sc v} also appear to be present, although the detection of the latter is not statistically significant and vanishes as the system evolves to lower X-ray luminosity. This pattern of line strengths -- especially the weakness of C\,{\sc iv} relative to  N\,{\sc v} -- is unusual for spectra formed in solar abundance material \citep[e.g.][]{Mauche1997ApJ...477..832M}. The implications of these line ratios for the evolutionary state of the system will be discussed in Section~\ref{sec:CNO}.

\begin{figure}
      \centering
      \includegraphics[width=0.48\textwidth]{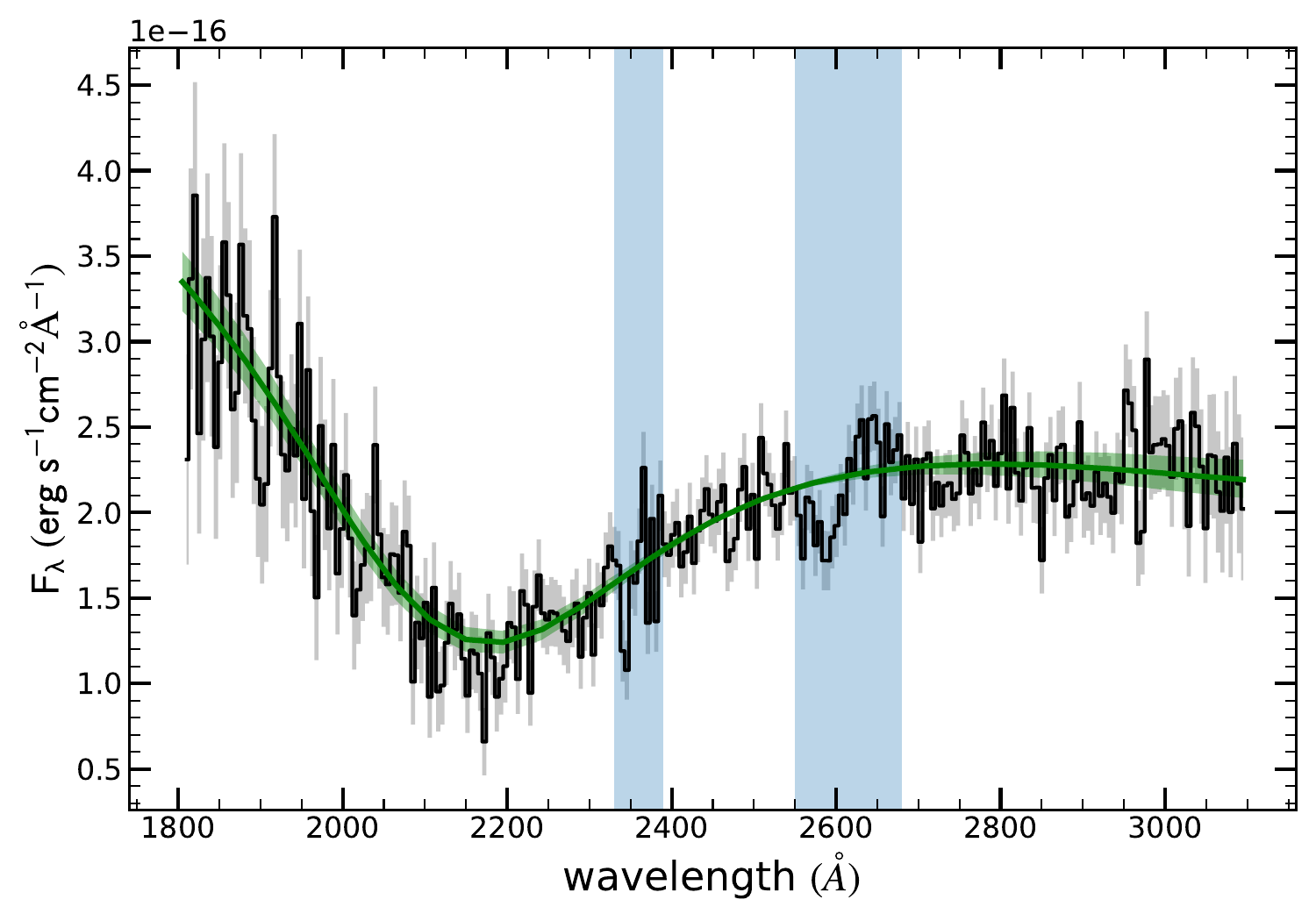}
      \caption[Extinction curve]{The near-UV spectrum of \src\ as seen by HST/STIS (black). The green line shows a reddened power-law fit to the data; the uncertainty around the fit is indicated as a shaded region. During the fit, the vertical shaded regions were masked to avoid Fe~{\sc ii} lines that might be present in the data. The Mg~{\sc ii} 2800 \AA\, line is not present in the observed spectrum.
      }
      \label{fig: E(B-V)}
    \end{figure}

 The strength of the interstellar absorption feature near $2175$\,\AA\ in the near-UV spectrum can be used to determine the the reddening, $E(B-V)$, towards \src\ ( see Fig.~\ref{fig: E(B-V)}). We estimate $E(B-V) \simeq 0.324 \pm 0.025$ from our data, corresponding to $A_V = R_V \, E(B-V) \simeq 1.00$ for $R_V = 3.1$ \citep{Cardelli1989ApJ...345..245C}. We adopt these values throughout this paper. Our reddening estimate for \src\ also implies a Hydrogen column density of $N_H \simeq 2\times 10^{21} {\rm\ cm^{-2}}$ \citep{Tolga2009_NH_to_EB-V}, consistent with the value obtained from its X-ray spectrum \citep{1858X-rayBursts2020MNRAS.499..793B}.

\subsection{Optical photometry} \label{sec:PANSTARRS}
In order to fully characterise the spectral energy distribution of \src\ in quiescence, we have collated  additional photometric measurements from PanSTARRS DR1 \citep[optical: grizy;][]{PANSTARR_DR1}. The PanSTARRS observations were carried out between MJD=55330 and 56830, a time when \src\ was in quiescence. 

\begin{figure}
      \centering
      \includegraphics[width=0.48\textwidth]{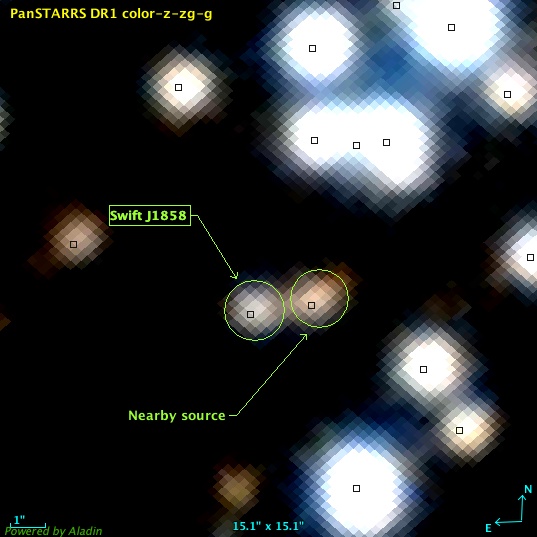}
      \caption[Field around \lmxb as seen in PanSTARRS]{Field around \lmxb~as seen in the archival images of PanSTARRS DR1. The quiescent counterpart of the XRB is clearly detected in this survey. A nearby source with centroid at $\sim 1$ arcsec apart is also indicated. The squares are the centroid of sources identified by {\it GAIA}. 
      }
      \label{fig: finding chart}
    \end{figure}

As can be seen in Figure \ref{fig: finding chart}, the field around \src\ is relatively crowded. In particular, there is a nearby source $\sim 1''$ West from the LMXB's optical counterpart. By default, the magnitude provided in PanSTARRS DR1 for each source in each band is generally an average over multiple frames. However, in order to avoid systematic errors due to blending with the nearby source -- especially in data taken under poor seeing conditions -- we retrieved the PSF-fitting measurements for each frame in each band using the {\it MAST} API\footnote{\url{https://catalogs.mast.stsci.edu/api/v0.1/panstarrs}}. We then calculated the median position of all the detections within $0.3 ''$ from the source and rejected all measurements that were offset more than $0.175 ''$ from this position or whose PSF was broader than $0.72 ''$ FWHM. This filtering yielded 2, 4, 4, 5, and 3 clean detections in the {\em grizy} bands, respectively, with corresponding flux densities: $g' = 2.1 (\pm 2),~ r' = 2.8 (\pm 1),~ i' = 2.3 (\pm 1),~ z' = 2.2 (\pm 2),~ y' = 1.9 (\pm 3)$ all in $10^{-17}\, erg\, s^{-1}\, cm{^2}\, $\AA$^{-1}$ . There is no statistically significant variability between the measurements in each band. The variability expected from ellipsoidal modulation for a Roche lobe  filling donor is comparable to the uncertainty, furthermore, most of the data points are obtained at similar amplitude of the modulation. therefore, we use the mean and error on the mean to estimate the quiescent flux and associated uncertainty for each filter.

\subsection{Optical spectroscopy}
\label{sec:SALT}

  We also obtained optical spectroscopy of \src\ in quiescence. The purpose of these observations was to check for the presence of strong emission lines in the quiescent optical spectrum. Such lines might point to a significant contribution of the accretion disk to the quiescent optical SED, and they might also themselves affect photometric measurements (e.g. H$\alpha$ lies within the $r$-band).
  
  These observations were carried out with the Southern African Large Telescope (SALT; \citealt{Buckley2006}) on 22 August 2022, approximately two years after the end of the outburst. 

  Two consecutive 1600~sec low-resolution spectra of \src\ were obtained in clear conditions.  The Robert Stobie Spectrograph (RSS; \citealt{Burgh2003}) was utilized in long-slit (1.25 arcsec wide) mode with the PG300 grating, covering the spectral range 4200-7260\AA~at a spectral resolution of 5.7\AA.
  The data were reduced using {\tt PySALT} version 0.47, the PyRAF-based software package for SALT data reductions \citep{Crawford2010}. These reduction tasks include cross-talk, bias, gain and cosmetic corrections. The spectral reductions (object extraction, wavelength calibration and background subtraction) were all carried out using standard IRAF tasks, with the wavelength calibration being performed using a Xe arc lamp exposure that was taken immediately after the observation. 
  There is no evidence of H$\alpha$ or any other emission lines. For H$_\alpha$, which we expect to be the strongest line, assuming an unresolved emission line (FWHM $\lesssim 6\,$\AA), the $ 3\sigma$ upper limit on the integrated line flux is
  $\simeq 2.6 \times 10^{-17}~\mathrm{erg\,cm^{-2}\,s^{-1}}$. This -- and the absence of variability in the PanSTARRS photometry -- suggests that the quiescent optical emission of \src\ is dominated by the donor star.
  
\begin{table}
    \centering
    \begin{tabular}{c|c|c|c|c|c|c}
         Visit & Line &  Flux& Err & 1$^{st}$moment & 2$^{nd}$moment\\
            &   &  $\rm cgs$ &  $\rm cgs$ & \AA & $\rm (km\ s^{-1})^2$\\
         \hline
         \hline
        1$^{st}$& N {\sc v} & 2.64e-13 & 5e-15& 1243.9 & 425.6\\
        & Si {\sc iv} & 5.6e-14 & 4.1e-15 & 1404.3 & 415.0\\
        & C {\sc iv} & <5.7e-15 & -- & -- & --\\
        & He {\sc ii} & 6.9e-14 & 4.5e-15 & 1641.7  & 254.2 \\
        \hline
        2$^{nd}$& N {\sc v} & 7.85e-14  & 1.27e-15 & 1242.3 & 469.0\\
        & Si {\sc iv} & 2.1e-14 & 2.0e-15 & 1402.0 & 597.0\\
        & C {\sc iv} & <1.8-15 & -- & -- & --\\
        & He {\sc ii} & 1.4e-14 & 1.5e-15 & 1640.9  & 468.0\\
        \hline
        3$^{rd}$& N {\sc v} &5.11e-14 & 7.7-16& 1241.7 & 538.2\\
        & Si {\sc iv} & 1.78e-14 & 7.4e-16 & 1401.5 & 967.4\\
        & C {\sc iv} & <7.5e-16 & -- & -- & --  \\
        & He {\sc ii} & 1.3e-14 & 1.0e-15 & 1639.2  & 628.4\\
        \hline
        4$^{th}$& N {\sc v} &1.20e-14 & 4.0e-16& 1241.7 & 552.6\\
        & Si {\sc iv} & 4.2e-15 & 5.3e-16 & 1392.2 & 931.1\\
        & C {\sc iv} & 7.8e-16 & 1.8e-16 & 1547.2 & 743.1\\
        & He {\sc ii} & 2.8e-15 & 3.0e-16 & 1641.1  & 529.9\\
        \hline
        \hline
         Q & $\rm H_\alpha$ & $ <2.6\times 10^{-17} $ & -- & -- & --  \\
        \hline
        \hline
    \end{tabular}
    \caption{Line fluxes in $\rm erg\ s^{-1}\ cm^{-2}$, estimated from the observed spectra after correcting for extinction. Far-UV lines are estimated from the four visits (labelled $1^{st}-4^{th}$ respectively, the corresponding spectra are illustrated in Figure~\ref{fig: fuv spectra}, covering the Luminous High-hard, Soft, Low-hard state (steady) and Low-hard states respectively, the first and second moment of the flux are also reported, i.e. the mean and variance of the distribution. The $ 3\sigma$ upper limit in {$\rm H_\alpha$} estimated during quiescent obtained two years after the end of the outburst, this visit is labelled as  Q.}
    \label{table: lines}
\end{table}

\subsection{Near-UV photometry}

  Another constraint on the relative contributions of the accretion disk and the donor to the optical SED can be obtained from near-UV photometry obtained by the Ultra-Violet/Optical Telescope (UVOT, \citet{uvot}) onboard the {\it Neil Gehrels Swift Observatory} \citet{swift}. 
  For each of UVOT's near-UV bands ($UVW2$, $UVM2$, $UVW1$ and $U$), we therefore combined all of the images obtained after the outburst ($MJD \geq 58940$). At this point, \src\ was no longer detectable in any of the individual near-UV images. We optimally stacked the quiescent images for each band and performed an optimal flux extraction by using {\sc TUVOpipe} \citep{TUVOpipe2022A&A...663A...5M}. This resulted in a solid detection for most of the bands: $UVW2_{\rm AB}=25.3\pm0.3\,{\rm mag}$, $UVM2_{\rm AB}>24.9\ (3\sigma),{\rm mag}$, $UVW1_{\rm AB}=25\pm0.3$ and $U_{\rm AB}=22.8\pm0.2\,{\rm mag}$ in the AB system. In Section~\ref{sec:PANSTARRS}, we combine these measurements with the PanSTARRS photometry to isolate the disk and donor contributions to the quiescent SED. 
  
\section{The nature of the donor star} \label{sec: results}

  
    The quiescent optical and near-IR emission of LMXBs with ``early'' type donors is often dominated by the companion star \citep[e.g. ][]{Torres2014MNRAS.440..365T}. As noted above, the lack of variability in the PanSTARRS photometry (Section~\ref{sec:PANSTARRS}) and the absence of emission lines in our SALT spectrum (Section~\ref{sec:SALT}) suggest this is also the case for \src .

    The quiescent SED of \src\ constructed from the Swift/UVOT and PanSTARRS photometry is shown in Figure~\ref{fig: companion SED}. It immediately suggests that the donor does indeed dominate the optical emission, with the disk only contributing significantly in the bluest near-UV bands.
    
    In order to quantify these statements and determine the parameters of the donor star, we have modelled the pre-outburst SED from PanSTARRS using synthetic models of stellar atmospheres. In doing so, we have also tested the effect of a quiescent accretion disk on the observed SED by including a fiducial disk contribution in our modelling. Both components and the modelling procedure are described in the following sections.

\subsection{The donor star SED model}

      We describe the donor SED as a simple model stellar atmosphere. More specifically, we use the grid of PHOENIX models \citep{phoenix} implemented in {\sc pysynphot}\footnote{https://pysynphot.readthedocs.io/en/latest/} for this purpose. The models on this grid are described by three parameters: (i) effective temperature, $T_{\rm eff}$; (ii) surface gravity, $\log{g}$; (iii) metallicity, Fe/H. When fitting these stellar SEDs to data, a fourth parameter is needed: (iv) the normalization, which is proportional to $R^2/d^2$. Here, $R$ is the donor radius, and $d$ is the source distance.

    \subsubsection{The accretion disk SED model}
    
    We model the accretion disk as a collection of concentric circular annuli, each of which is characterized by an effective temperature $T_{\rm eff}(R)$. The disk extends from the surface of the NS, $R = R_{\rm NS}$ (the exact value may be bigger, but ths part of the disk would radiate only X-rays), to an outer radius $R = R_{\rm disk}$. The effective temperature is set by the requirement that the rate at which an annulus radiates energy away,  
    $\sigma T_{\rm eff}^4$, must balance the rate at which energy is deposited into it by viscous dissipation and irradiation. The viscous heating rate can be written as \citep[e.g.][]{FrankKingRaine2002apa..book.....F}, 
    \begin{equation}
        \sigma T_{\rm visc}^4 = \frac{3 G M_{\rm NS} \dot{M}_{\rm acc}}{8\pi R^3} \left[ 1 - \left(\frac{R_{\rm NS}}{R}\right)^{1/2}\right],
    \end{equation}
    where $M_{\rm NS}$ is the mass of the neutron star, and $\dot{M}_{\rm acc}$ is the accretion rate. The heating rate due to irradiation can be modelled as
     \begin{equation}
        \sigma T_{\rm irr}^4 = \left(\frac{L_{\rm irr}}{4\pi R^2}\right) 
        \left(\frac{H}{R}\right) \gamma 
        \left(1-A\right),
    \end{equation}
    where $L_{\rm irr}$ is the irradiating luminosity (assumed to originate from a central point source), and $A$ is the albedo (so that $1-A$ is the fraction of the light incident on the annulus that is absorbed). The quantity $H/R$ is the aspect ratio of the disk, which can be shown to scale as 
    \begin{equation}
        \frac{H}{R} = \left(\frac{H}{R}\right)_{R_{\rm disk}} \left(\frac{R}{R_{\rm disk}}\right)^\gamma, 
    \end{equation}
    where $\gamma = 1/8$ in the absence of irradiation and $\gamma = 2/7$ if irradiation dominates the heating rate. Strictly speaking, $\gamma$ is therefore a function of radius, but we neglect this here and simply set $\gamma = 2/7$ everywhere. This approximation means that we will slightly overestimate the influence of irradiation, but only in disk regions where irradiation is relatively unimportant anyway. Putting all of this together, the effective temperature of the disk can be calculated by requiring that total heating should be matched by radiative cooling, i.e.
    \begin{equation}
        \sigma T_{\rm eff}^4 = \sigma T_{\rm visc}^4 + \sigma T_{\rm irr}^4.
    \end{equation}

    In order to calculate the spectrum of the disk, we assume that each annulus radiates as a modified blackbody, 
    \begin{equation}
        B_{\nu, {\rm mod}}(f,T_{\rm eff}) = \frac{2 h \nu^3}{f^4 c^2 
    \left[e^{\frac{h\nu}{f k T_{\rm eff}}} - 1\right]}.
    \end{equation}
    Here, $f$ is the so-called "spectral hardening factor" \citep{ShimuraTakahara1995ApJ...440..610S}, which approximately corrects for the effects of Compton scattering in the disk atmosphere. This factor is not actually a constant, but rather a function of temperature, surface density and radius (and, for black holes, spin parameter). In our SED model, we parameterize $f$ using the analytical fitting function provided by \citet[][their Equation~10]{DavisEl-Abd2019ApJ...874...23D}. In calculating the required surface density, we take into account the relevant relativistic correction factors \citep{DavisEl-Abd2019ApJ...874...23D,NovikovThorne1973blho.conf..343N}.

    \subsubsection{SED modelling} \label{sec: SED}

\begin{figure}
      \centering
      \includegraphics[width=0.499\textwidth]{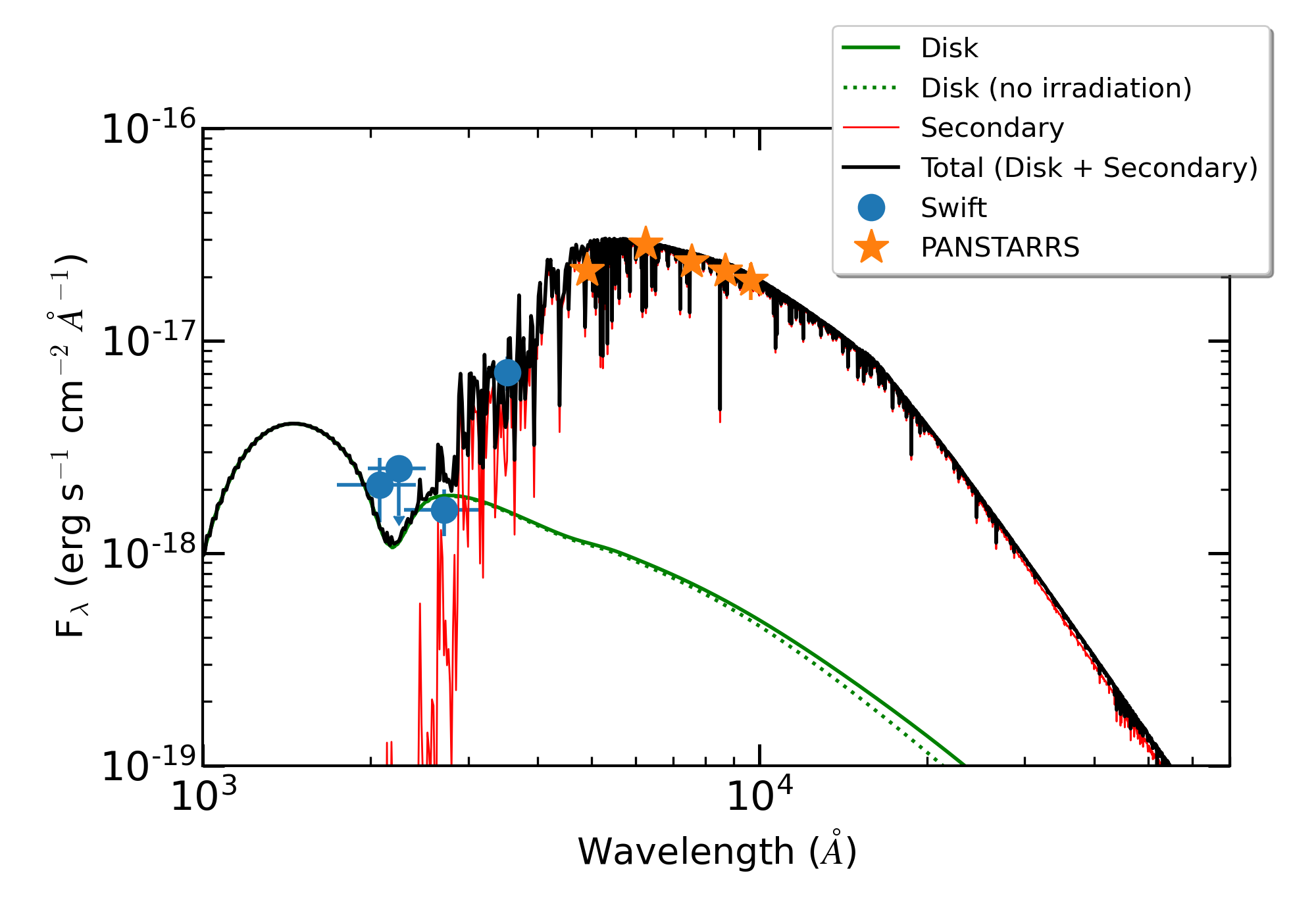}
      \caption[Quiescent SED of \lmxb~in different bands]{Archival quiescence SED of \lmxb~in different bands (stars), and near-UV measurements during quiescent after the 2018-2020 outburst (blue circles), see text for details. A {\rm $T_{\rm eff}= 5700\,K$} stellar model is shown (red line), with a fiducial irradiated accretion disk model (green line). The same disk spectrum with no irradiation is also shown (dotted green line).
      }
      \label{fig: companion SED}
    \end{figure}

    We are mainly interested in the properties of the donor star. We therefore restrict our modelling to the PanSTARRS bands, since the observed SED in Figure~\ref{fig: companion SED} already shows that the accretion disk contributes little to the quiescent flux in this spectral region. As a precaution, we nevertheless include a "reference" disk model SED in our modelling and then test {\it a posteriori} whether reasonable changes in the disk parameters affect the inferred donor properties.
    
    The most important disk parameter is the accretion rate. In our reference model, we fix this by requiring that the disk SED should roughly match the Swift/UVOT near-UV fluxes. The implied post-outburst accretion rate is $\dot{M}_{\rm acc} \simeq 10^{-9.5}\ {\rm M_\odot\ yr^{-1}}$, in line with \src\ being a transient source \citep[e.g.][]{Coriat+2012MNRAS.424.1991C,Dubus2019A&A...632A..40D}. Other parameters of the reference model (e.g. disk radius, albedo, H/R) are set to plausible values derived from modelling the SED during outburst (details of this modelling effort will be presented in a separate paper). 

    We then model the PanSTARRS photometry as a sum of reddened (Section~\ref{sec: uv_spec}) disk + donor SEDs. In carrying out the necessary synthetic photometry, we adopt the transmission curves obtained from the SVO filter service\footnote{http://svo2.cab.inta-csic.es/theory/fps/} for the PanSTARRS bands. Since the disk model is treated as fixed, the only free parameters are those describing the donor star, i.e. normalization ($R^2_{\rm donor}/d^2$), $T_{\rm eff}$, $\log{g}$, and Fe/H. Unsurprisingly, we find that the latter two parameters are essentially unconstrained by the photometric data. The uncertainty associated with reddening is accounted for by including $E(B-V)$ as a hyperparameter with a Gaussian prior (see Section~\ref{sec: uv_spec}). For completeness, we have also carried out the fitting of the donor SED with no disk.

    Our best-fitting model is shown in Figure \ref{fig: companion SED}. The implied donor temperature is $T_{\rm eff} = 5700 \pm 300 {\rm ~K}$, while the normalization is $\log_{10}(R^2_{\rm donor}/d^2) = -23.05 \pm 0.08$.  As expected, these parameters turn out to be insensitive to any reasonable changes in the description of the accretion disk, so long as the SED remains consistent with the near-UV constraints. We note that UVOT's PSF may introduce some blending from the nearby source, and the near-UV filters also suffer from non-negligible red leaks. If these effects are significant, they would imply an even lower near-UV contribution of the accretion disk in quiescence. To test the possible impact of this on our measurements, we repeated the fit after removing the accretion disk contribution entirely from the SED modelling. We obtained fit parameters well within the statistical uncertainties quoted above.
    

    In order to determine the radius and luminosity of the donor from the normalisation of the model fit we require a distance estimate. \citet{1858X-rayBursts2020MNRAS.499..793B} have obtained this by modelling 
    the evolution of the Type I X-ray bursts in this system in their time-resolved X-ray spectroscopy. Briefly, when the bursts are triggered, the luminosity at the surface of the star reaches the Eddington limit. At this point, radiation pressure lifts the material from the surface. The luminosity is expected to remain (approximately) constant at $L_{Edd}$ during this expansion phase and the subsequent contraction of the atmosphere. Thus modelling of these bursts can be used to estimate the distance to the source \citep{vanParadijs1978Natur.274..650V,Kuulkers2003AA...399..663K}. Here, we adopt the most likely value suggested by \citet{1858X-rayBursts2020MNRAS.499..793B}, $d = 12.8\pm 0.7\ {\rm kpc}$. Systematic uncertainties associated with reddening (e.g. the choice of the reddening law) might introduce additional errors on the order of $\sim 20\%$. 
    
\begin{figure}
      \centering
      \includegraphics[width=0.49\textwidth]{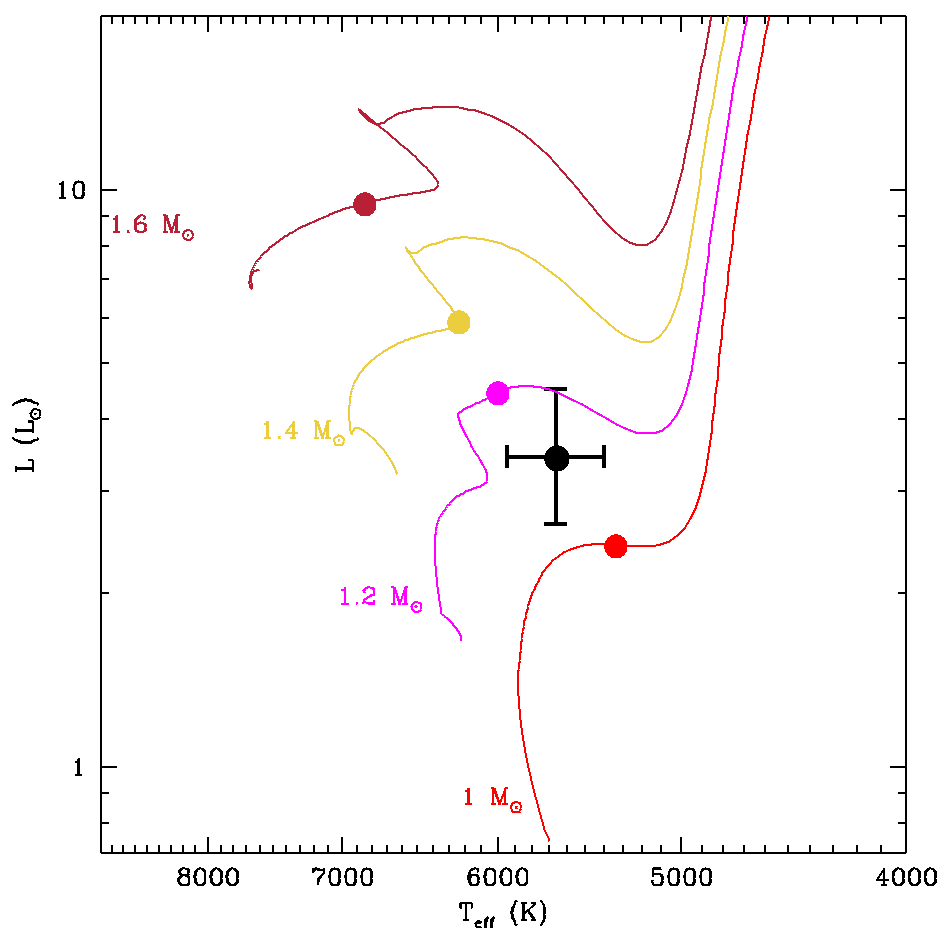}
      \caption[Hertzsprung–Russell diagram of the quiescent counterpart ]{HRD position of \src\ is shown as a black cross, with evolutionary tracks for isolated stars of different masses plotted as solid lines.  The coloured dots show the positions where these stars would fill their Roche Lobe given the 21.3 hr $P_{\rm orb}$.  The current position of \src\ in the HRD is  consistent with a fairly massive star ($\approx 1.1 M_\odot$), emerging from the main sequence.
      }
      \label{fig: HRD}
    \end{figure}

    At this distance, the implied donor radius is $R_2 \simeq 1.7 \pm 0.15 R_\odot$, with a corresponding luminosity of $L_2\simeq 3.5 \pm 0.9 L_\odot$. We can therefore now place the donor in a Hertzsprung-Russell diagram (HRD), as shown in Figure~\ref{fig: HRD}. The position of \src's quiescent counterpart in the HRD is shown as a black cross, overlaid on top of a set of evolutionary tracks of isolated stars with masses ranging from $1$ to $1.6 M_\odot$ computed with {\tt MESA}\footnote{http://mesa.sourceforge.net} \citep{MESA}. The corresponding radius at which the isolated stars fill their Roche Lobe for the current orbital period of \src\ is marked with a coloured circle on the tracks. The HRD shows that the donor currently resembles a $\simeq 1.15 {\rm M_\odot}$ sub-giant that is evolving from the terminal main sequence into the Hertzsprung gap.

\section{Discussion}\label{sec: discussion}

   \subsection{Evidence for CNO processing}\label{sec:CNO}

    \begin{figure*}
      \centering
      \includegraphics[width=0.98\textwidth]{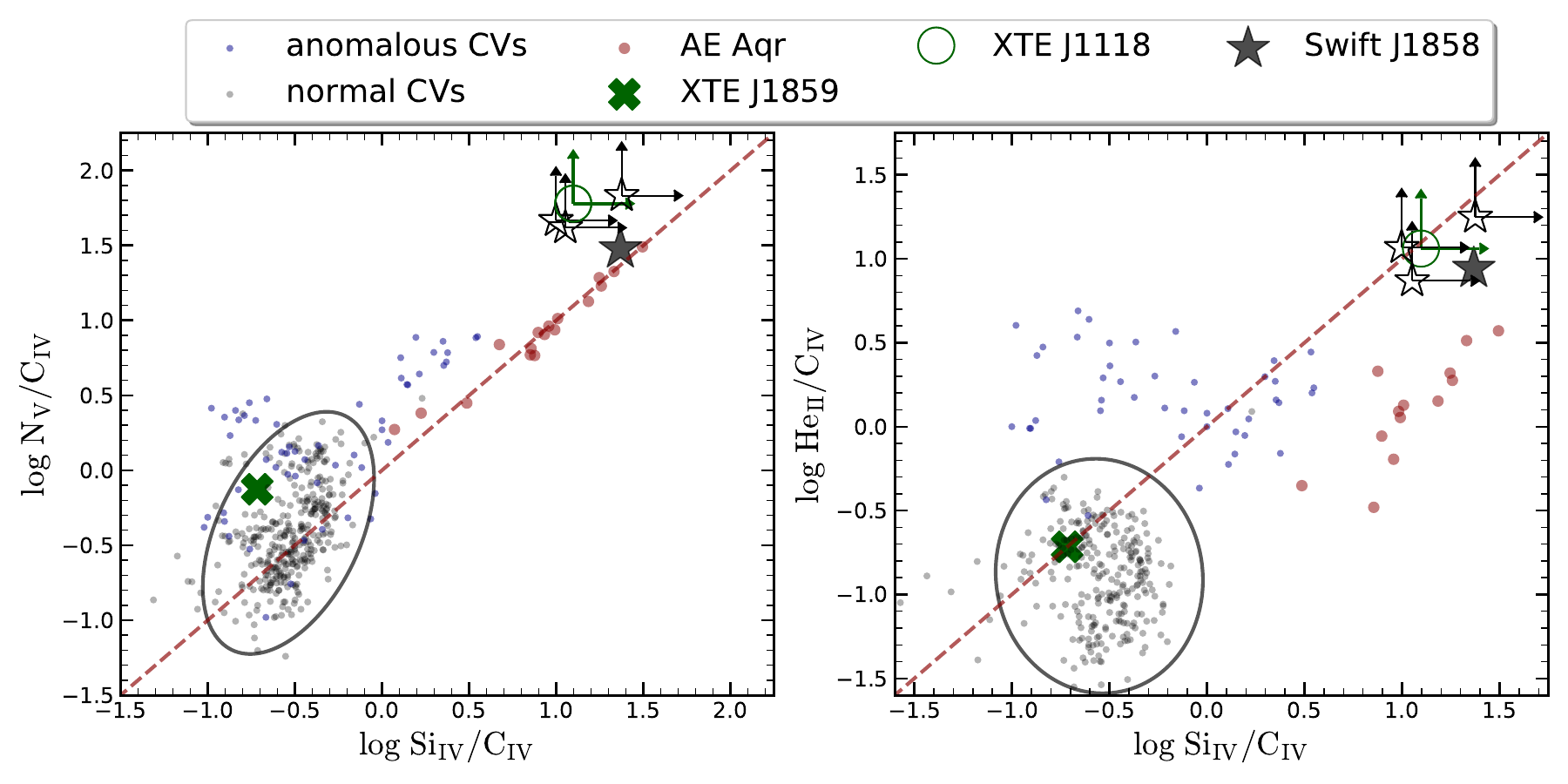}
      \caption[Far-UV line flux ratios for CVs (dots) and LMXBs]{Far ultraviolet line flux ratios for CVs (dots) from \citet{Gansicke2003ApJ...594..443G}. {\it XTE J1118+480} and {\it XTE J1859+226} LMXBs (open circle and filled cross respectively),  from \citet{Haswell2002MNRAS.332..928H}. While most CVs accrete from near solar-abundances (encircled grey dots), a set of CVs display anomalous line ratios (blue dots), which as believed to be a consequence of a enhanced CNO-processed material from the donor. {\it XTE J1118+480}  and {\it AE Aqr} display the most extreme ratios, thought to be a consequence of higher initial secondary masses, $M_{2,i}\gtrsim 1.5$.  \src\ (black stars) has very similar ratios to {\it XTE J1118+480}.  Upper limits are shown as non-filled symbols. Note that the flux level of C~{\sc iv} in \src\ is marginal, and variations on the selection of continuum region could lead to a upper limit in the flux, while for {\it XTE J1859+226}, the errors are smaller than the symbol. The typical errors for CVs are $\sigma\sim0.2\,{\rm dex}$ \citep[see][ for an object by object detailed analysis]{Mauche1997ApJ...477..832M}.
      }
      \label{fig: line ratios}
    \end{figure*}           
    The far-UV range covered by HST/COS contains several spectral lines whose ratios can be used as indicators of CNO processing \citep[e.g. ][]{Mauche1997ApJ...477..832M,Haswell2002MNRAS.332..928H,Gansicke2003ApJ...594..443G,Froning2011ApJ...743...26F,Froning2014ApJ...780...48F}. For example, Nitrogen is strongly enhanced by the action of the CNO cycle, while Carbon is suppressed. As a result, the ratio (N\,{\sc v}$\rm\lambda$1240$/$C\,{\sc iv}$\rm\lambda$1550\,) is expected to be much higher if the lines are formed in CNO-processed material. When such anomalous line ratios are detected in compact binaries, they suggest that the accreting material has undergone CNO processing. This immediately implies that the initial donor mass must have been $\gtrsim 1.4 M_\odot$, since the CNO cycle only becomes important at the core temperatures found in stars above this mass limit \citep{Clayton1983psen.book.....C}.

    The key UV line ratios for \src\ are presented in Figure~\ref{fig: line ratios}, along with those of two other LMXBs and cataclysmic variables (CVs; compact binaries in which the accretor is a white dwarf). Most of the CVs in this figure (black dots) harbour a (roughly) main-sequence donor star and have evolved via the standard evolutionary channel for CVs. In systems following this channel, the donor star is essentially unevolved and mass-transfer is driven by angular momentum losses \citep[AML; e.g.][]{Knigge2006}. However, several CVs clearly exhibit anomalous line ratios (blue dots). These are thought to be those in which the initial mass of the present-day donor was substantially higher. Here, contact can be initiated as a result of the donor's expansion due to nuclear evolution at a mass ratio $q = M_{\rm WD}/M_{\rm donor} \geq 1$. This results in a short phase of unstable thermal time-scale mass transfer \citep[TTSMT; e.g.,][]{ Paczynski1969ASSL...13..237P} during which the donor loses a significant part of its envelope. It therefore emerges from this phase with a composition (and surface abundances) that more closely resembles the CNO-processed material in its stellar core \citep{SchenkerKing2002ASPC..261..242S}. Once the mass ratio has been reduced to $q \lesssim 1$, these systems can become ``normal'' (AML-driven) accreting binaries. However, they will now exhibit "anomalous" line ratios in their far-UV spectra \citep[blue dots in Figure~\ref{fig: line ratios}; ][]{Mauche1997ApJ...477..832M,SchenkerKing2002ASPC..261..242S}. The famous magnetic propeller system {\it AE Aqr} (grey circles) is the most extreme of these anomalous CVs and is believed to have emerged from its super-soft X-ray binary phase relatively recently \citep{Schenker_AE_Aqr_2002MNRAS.337.1105S}. 

    The two other LMXBs shown included in Figure~\ref{fig: line ratios} are {\it XTE J1859+226} and {\it XTE J1118+480}, which have been discussed in \citet{Haswell2002MNRAS.332..928H}. Along with \src\, these three LMXBs cover the same range of line ratios found among the CV population. More specifically, {\it XTE J1859+226} exhibits line ratios typical of normal CVs, while {\it XTE J1118+480} and \src\ both display anomalous ratios. {\it XTE J1118+480}, is a short-period system ($\rm P_{\rm orb}\simeq 4\ h$) system in which a $\sim 6 M_\odot$ BH is thought to accrete CNO-processed  material from its stripped donor \citep{Haswell2002MNRAS.332..928H}. The pattern of line ratios in \lmxb\ is one of the most extreme among all the compact binaries in Figure~\ref{fig: line ratios}, rivalled only by {\it XTE J1118+480} and AE\,Aqr.

    \begin{figure*}
      \centering
      \includegraphics[width=0.98\textwidth]{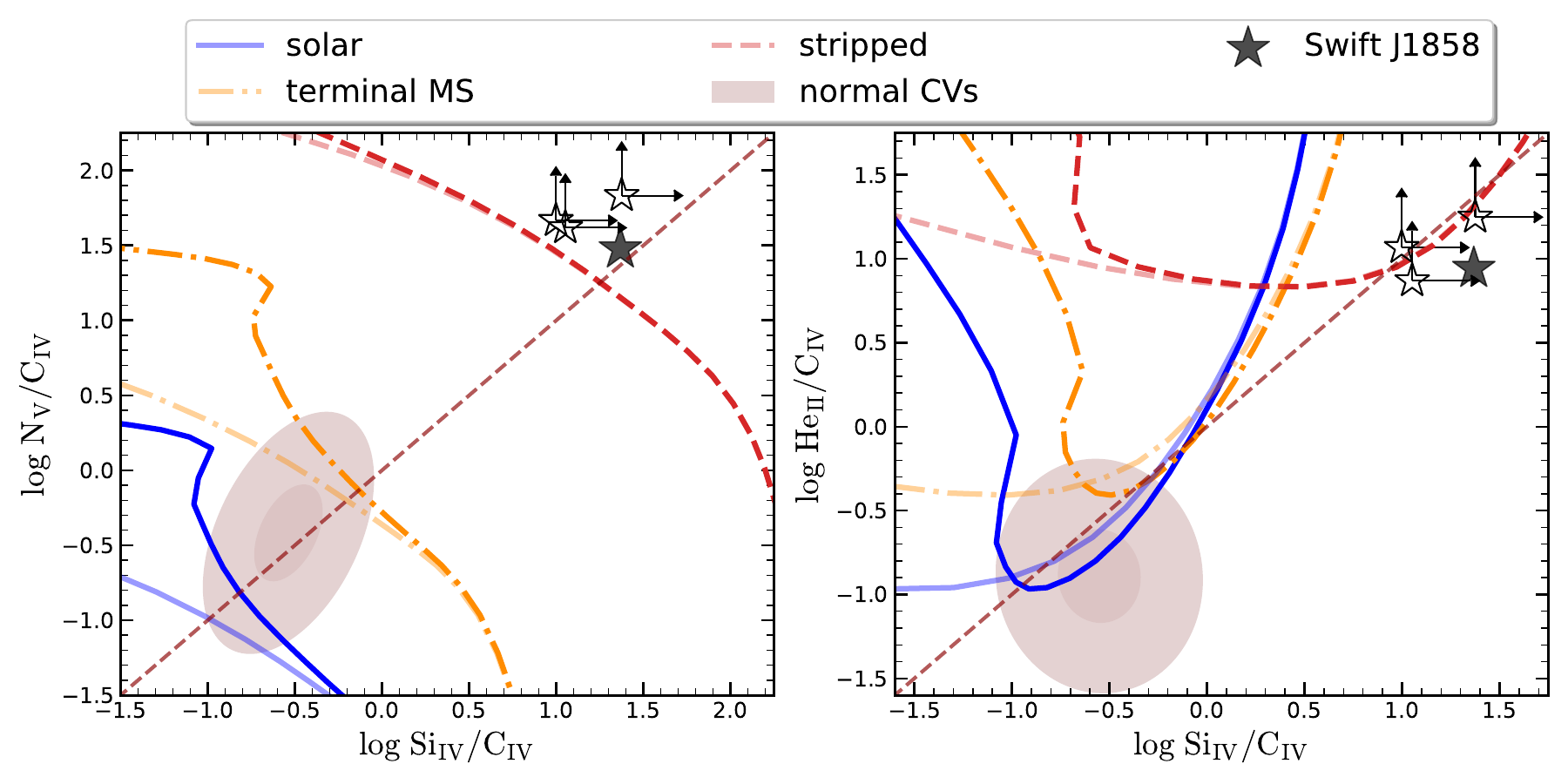}
      \caption[]{Same as Figure~\ref{fig: line ratios}. Here, the population of ``normal'' CVs is shown for reference as the shaded regions with semi-major axis as $1\sigma$ and $2.5 \sigma$ of the population. The lines are predicted line ratios as a function of the ionization parameter computed with {\sc cloudy} for an optically thin parcel of gas irradiated with a simple accretion disc. Solid, dash-dot and dashed lines are models carried out with solar abundances, $M\approx1.5-2M_\odot$ terminal main sequence and the equilibrium CNO-cycle core of a $M\approx2M_\odot$ star respectively. The latter is labelled as stripped, as we consider to have representative abundances of a stripped star with its convective CNO core exposed. The models include the O\,{\sc iv} multiplet within range of the Si\,{\sc iv} doublet, for reference lines including the emission only from Si\,{\sc iv} are shown with the same linesytle but higher transparency. The measurements of \src\ (filed and non-filled black stars), lie very close to the stripped model.} 
    
      \label{fig: line ratio models}
    \end{figure*}           

    Although details of the UV line formation mechanism(s) in accreting binaries are uncertain,  it is instructive to compare the observed line ratios to those predicted by simple toy models. Following \citet{Mauche1997ApJ...477..832M}, we model the line emission as arising in an optically thin cloud of gas that is irradiated and photo-ionized by a point source. For the SED, we adopt a simple accretion disk model (a multi-colour blackbody) with inner temperature $T_{in}= 2 keV$; this should be a reasonable first approximation for LMXBs. 
    
    For any given set of abundances, the line ratios in such a model will depend almost entirely on the ionization parameter, $\xi = L_{\rm ion} / (r^2 n_H)$. Here, $L_{\rm ion} \propto L_{\rm acc}$ is the ionizing luminosity, $r$ is the distance between the cloud and the irradiating source, and $n_H$ is the Hydrogen number density of the cloud. We therefore use {\sc cloudy} c17.2 \citep{cloudy17}, to predict the UV emission line ratios for this setup for a wide range of $\xi$ values. In principle, the density itself can also affect the line ratio (by favouring collisional over radiative de-excitation), but these effects are small compared to the differences between "normal" and "anomalous" ratios we are interested in here  \citep[c.f.][]{Temple2021MNRAS.505.3247T}. 
   
    For the abundances, we consider three different cases designed to correspond to (i) solar abundances; (ii) the globally averaged abundances in a $\simeq 2~M_{\odot}$ terminal-age main sequence star; (iii) the abundances in the stripped core of a $\simeq 2~M_{\odot}$ star. 
    For case (ii), we adopt $(X/X_\odot)_{^2He} \approx 2$, $(X/X_\odot)_{^6C} \approx 0.5$, $(X/X_\odot)_{^7N} \approx 7$ and $(X/X_\odot)_{^{16}O} \approx 1$ for Helium, Carbon, Nitrogen and Oxygen, respectively \citep{Gallegos-Garcia2018ApJ...857..109G}. For case (iii), we use the calculations by \citet{Gervino2005JPhG...31S1865G} to estimate the equilibrium abundances of Carbon, Nitrogen and Oxygen produced by the CNO cycle for the core temperature of a $\simeq 2~M_{\odot}$ main-sequence star ($T_{c} \simeq 2 \times 10^7$\,K). These estimates were  
     $(X/X_\odot)_{^6C} \approx 0.018$, $(X/X_\odot)_{^7N} \approx 7.5$ and $(X/X_\odot)_{^{16}O} \approx 0.037$. Note that Helium is a product, rather than a catalyst, for the CNO cycle, so it is not possible to estimate an equilibrium abundance for this element.

    As discussed extensively in \citet{Mauche1997ApJ...477..832M}, the comparison of models with observations is complicated by the possibility that observed emission lines could be  contaminated (or even dominated) by transitions other than the intended ones. In particular, the N\,{\sc v} doublet can be polluted by  the Mg\,{\sc ii} $^3S - ^4P$ doublet (at ${\rm \lambda\lambda 1238.82, 1242.80 }$\,\AA\ and ${\rm \lambda\lambda 1239.93, 1240.39 }$\,\AA, respectively). However, the Mg\,{\sc ii} lines only contribute significantly in very low ionization conditions, and the NUV spectrum of \lmxb\  does not exhibit any sign of the most prominent Mg\,{\sc ii} line, the resonance doublet near 2800\,\AA. We therefore assume that the observed emission line near 1240\,\AA\ is entirely due to N\,{\sc v}. 
    The Si\,{\sc iv} resonance doublet near 1400\,\AA\ can in principle also be contaminated, in this case by 
    an O\,{\sc iv} multiplet located between the doublet's components. Since the ionization states favourable to Si\,{\sc iv} and O\,{\sc iv} are fairly similar, in this case we include the fluxes from both lines in our model calculations. However, we present results for the ``isolated'' Si\,{\sc iv} line flux as well.  

    The resulting model tracks for the line ratios are shown in Figure~\ref{fig: line ratio models}. Overall, they support the idea that "normal" line ratios correspond to roughly solar-abundance accreting material, whereas ``anomalous'' line ratios are associated with accreting material that has undergone some degree of CNO processing \citep{Mauche1997ApJ...477..832M,Haswell2002MNRAS.332..928H,SchenkerKing2002ASPC..261..242S,Schenker_AE_Aqr_2002MNRAS.337.1105S,Gansicke2003ApJ...594..443G}. For \lmxb, in particular, Figure~\ref{fig: line ratio models} shows that only the case (iii) track comes close to matching the observed line ratios. As noted above, the abundances adopted for this track correspond to those in the stripped core of a 2~$M_{\odot}$ star where equilibrium CNO abundances have been reached. The evidence of CNO processed material being accreted in the compact object, suggest an initial donor mass was $M_d\approx 2 M_{\odot}$.

    \begin{figure}
      \centering
      \includegraphics[width=0.49\textwidth]{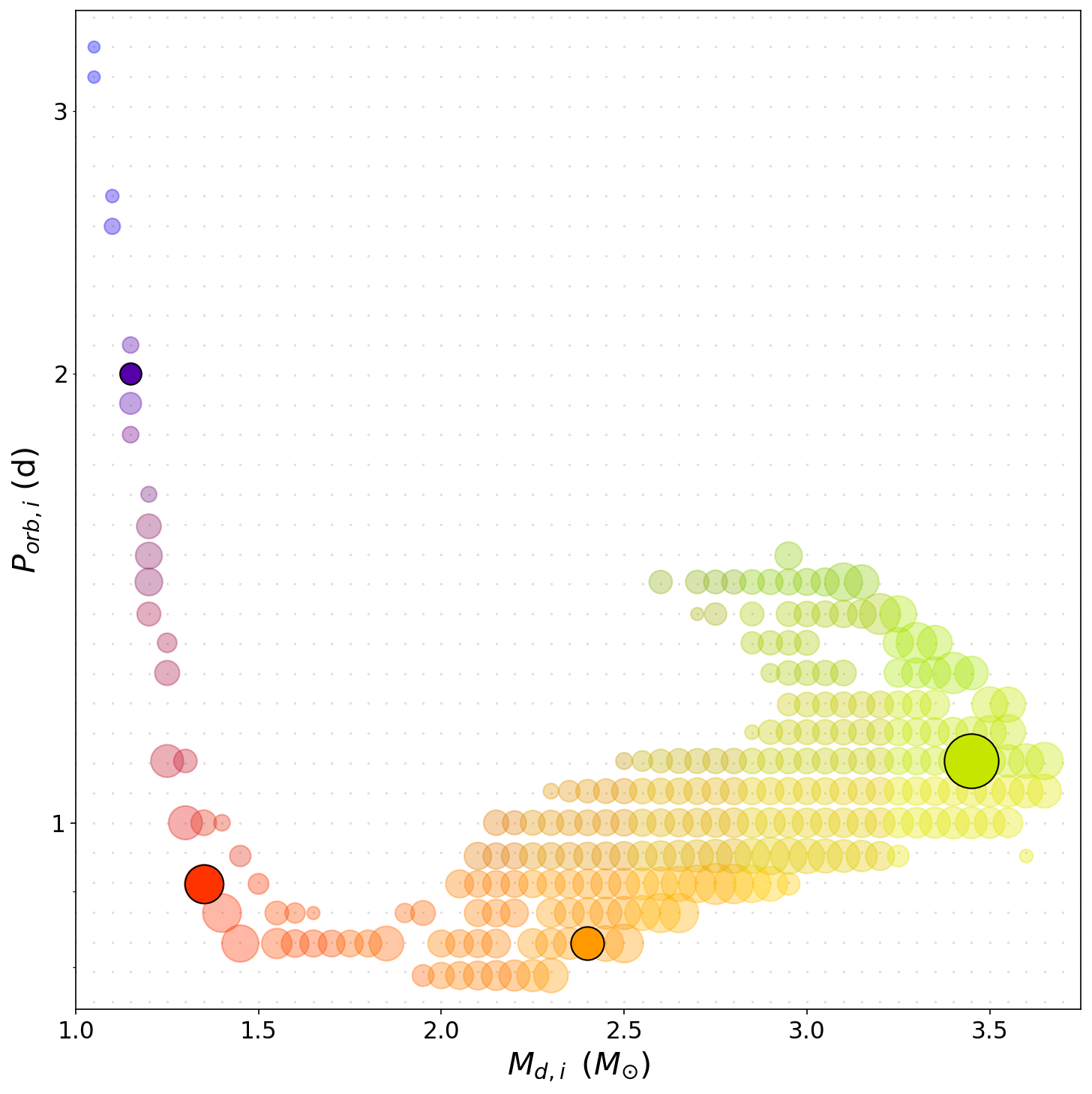}
      \caption[parameter space]{Initial orbital period vs initial donor mass parameter space for all the models compatible with the observed effective temperature and luminosity of \lmxb. The size of the circles is associated with the time expend by the binary model in the observed ($\simeq 21{\rm h}$) orbital period. The colour scale is representative of different combination of parameters leading to the distinct evolutionary scenarios (see text for details). The highlighted models are the reference examples of these scenarios and are be carried over the following figures.
      }
      \label{fig: param space}
    \end{figure}     

    \begin{figure}
      \centering
      \includegraphics[width=0.49\textwidth]{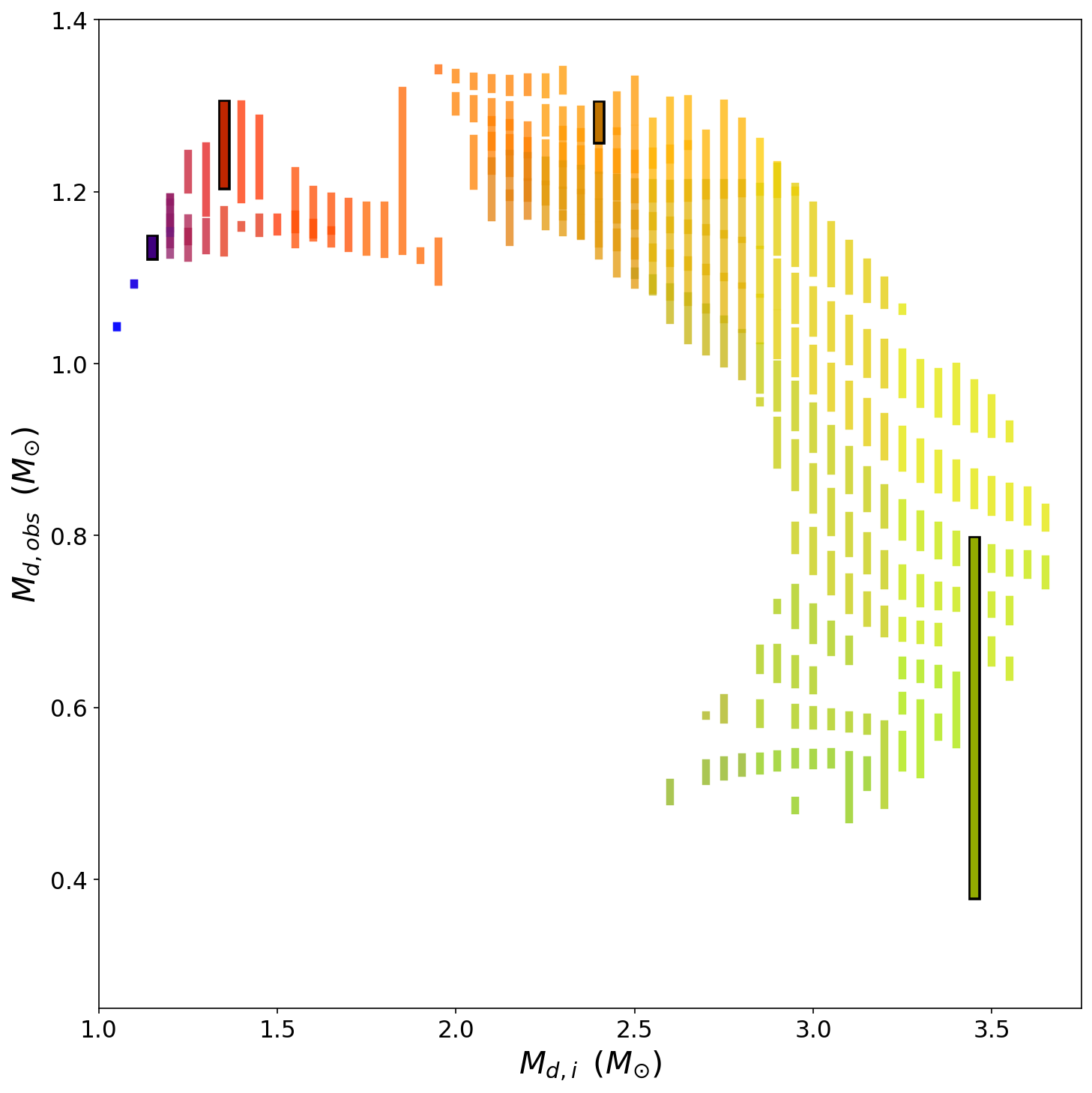}
      \caption[ ]{Observed mass ($M_{d,{\rm obs}}$) vs initial mass ($M_{d,{\rm i}}$), for all the models. The entire range of donor mass in the current evolutionary stage of \src~is shown for each of the allowed models. Allowed masses for \src\ at its current state are in the range $M_{d,{\rm obs}} \approx 0.5 -- 1.3 M_\odot$. The reference models in Fig.~\ref{fig: param space} are highlighted, colours are matched in both figures.  
      }
      \label{fig: Mobs vs Mi}
    \end{figure}

    \begin{figure*}
      \centering
      \includegraphics[width=0.98\textwidth]{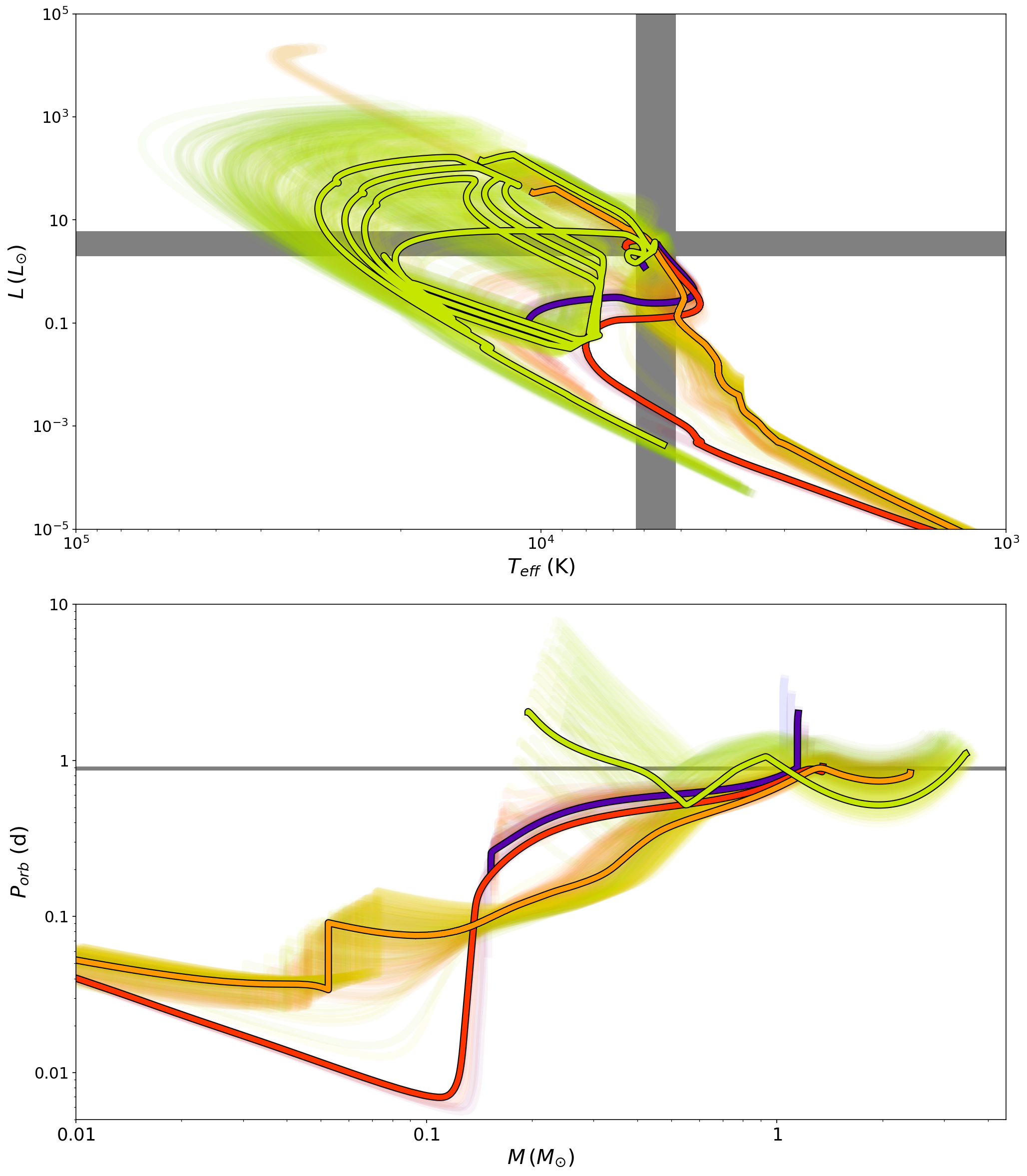}
      \caption[Evolutionary tracks for binary sequences in the $P_{\rm orb}$ vs $M_d$ and $L_d$ vs $T_{\rm eff,d}$ planes]{
      Evolutionary tracks for all the models compatible with the observed parameters of \src. Tracks for \src's compatible binary sequences in the $L_d$ vs $T_{eff,d}$ (top), and $P_{\rm orb}$ vs $M_d$ planes (bottom). Initial secondary masses and orbital period are drawn from the space parameter presented in Figure~\ref{fig: param space}. All the allowed solutions shown in Figs.~\ref{fig: param space}--\ref{fig: Mobs vs Mi} are also plotted with the same colour scale (transparent tracks). Highlighted models are the same as in Figs.~\ref{fig: param space}--\ref{fig: Mobs vs Mi}. The shaded regions indicate the allowed range in the parameter space (see text for details).
      }
      \label{fig: tracks}
    \end{figure*}

\subsection{Evolutionary tracks: past, present and future}\label{sec: evolution}

We have shown that \src\ displays anomalous far-UV emission line ratios that suggest the accreting material has undergone CNO processing. This, in turn, implies that the initial mass of the donor must have been $M_{d,i} \gtrsim 2~M_{\odot}$. We have also been able to estimate the radius $R_d \simeq 1.7~R_{\odot}$ and $T_{d} \simeq 5700$~K, of the donor, based on the quiescent SED of the system. Our goal now is to isolate the set of binary evolution tracks that is consistent with these observational constraints.

In order to set the scene, Figure~\ref{fig: HRD} shows the location of the donor star in a theoretical Hertzsprung-Russel diagram (HRD), alongside several theoretical evolution tracks for {\em single}, solar-abundance stars. This immediately shows that the donor is too red and luminous to be on the main sequence. Instead, its position on the HRD corresponds to that of a Roche-lobe-filling sub-giant star with $M \simeq 1.1~M_{\odot}$ \citep[c.f.][]{1858_Eclipses2021MNRAS.503.5600B}. This is significantly smaller than the initial donor mass implied by the evidence for CNO processing, suggesting that the donor has lost a significant fraction of its envelope via mass transfer to the NS primary. In fact, the present-day donor mass could be substantially less than suggested by single star evolution tracks, since heavily stripped stars tend to be over-luminous for their mass \citep{Giannone1968ZA.....68..107G,SchenkerKing2002ASPC..261..242S}.

In order to constrain the evolutionary state and path of \src\ more quantitatively, we have used the Database of Accreting Binary Simulations \citep[DABS;][]{DABS_Mangat2023A&C....4200681M}. This is an open-access database of theoretical binary evolution tracks for LMXBs, simulated with {\tt MESA} \citep{MESA}. All  tracks in DABS start from a binary configuration in which the compact object -- a $1.4~M_{\odot}$ NS or a BH with mass $5~M_{\odot}$, $7~M_{\odot}$ or $10~M_{\odot}$ -- has already formed. This removes the need to deal explicitly with the uncertain common envelope (CE) phase in the simulations. The binary orbit is assumed to be circular, with initial period $P_{{\rm orb},i}$. The companion starts as a zero-age, solar-abundance main sequence star with initial mass $M_{d,i}$. For each type of compact object, the evolution tracks in DABS cover the parameter space $0.95~M_{\odot} \leq M_{d,i} \leq 7.0~M_{\odot}$ and $0.25~\mathrm{d} \leq P_{{\rm orb},i} \leq 10^4~\mathrm{d}$. 

Aside from the CE phase, the most uncertain aspect of such binary evolution calculations is the treatment of magnetic braking \citep[MB; e.g. ][]{Rappaport1983ApJ...275..713R}. This term refers to the angular momentum loss associated with a magnetic wind from the companion star, which can be an important driver of binary evolution. The tracks in DABS use the Convection-and-Rotation-Boosted (CARB) recipe for MB \citep{CARB_MB_VanIvanova2019ApJ...886L..31V}.

An important tool provided with DABS is the "Progenitor Extractor for Accreting Systems" (PEAS). PEAS allows the user to specify constraints on the present-day properties of an LMXB and then searches DABs to identify those evolution tracks that pass through the allowed region of parameter space. The specific properties that can be matched by PEAS are the donor mass and effective temperature, the accretor type and mass, the orbital period and the mass-transfer rate. Since we additionally have a constraint on the donor radius, we have modified PEAS to allow  this property (or alternatively the luminosity of the donor) to be matched as well. 

In searching for viable progenitor systems for \src, we demand that the accretor is a NS, but treat both of the present-day component masses as unknown. The present-day orbital period is required to be within $\pm 0.5$~hrs of the actual $P_{\rm orb} \simeq 21.3$~hrs. This range is much larger than the observational uncertainties, but is adopted to account for the inevitable systematic uncertainties associated with theoretical binary evolution tracks (due to, for example, the adopted MB recipe). The present-day radius and effective temperature of the donor are required to fall within $1.5~\mathrm{M_{\odot}} \leq R_{d} \leq 2.0~\mathrm{M_{\odot}}$ and 
$5140~\mathrm{K} \leq T_{d} \leq 6240~\mathrm{K}$, respectively, roughly the 2-$\sigma$ range allowed by our SED modelling (see Sec.\ref{sec: SED}).

The final constraint we adopt is based on the transient nature of \src, which implies that the system is subject to the thermal-viscous disk instability \citep{Lasota2016}. The present-day mass-transfer therefore has to be less than the critical rate above which the instability is quenched. For a NS LMXB with $P_{\rm orb} = 21.3$~hrs, this requirement translates to $\dot{M}_{tr} \lesssim 10^{-8.2}~M_{\odot}~\mathrm{yr}^{-1}$ \citep{Coriat+2012MNRAS.424.1991C,Dubus2019A&A...632A..40D}.

The results from this selection are summarised in Figures~\ref{fig: param space}, \ref{fig: Mobs vs Mi} and \ref{fig: tracks}. Figure~\ref{fig: param space} shows the resulting initial values of the parameter space in the simulations after all the constraints have been applied. The allowed tracks cover $1.1~M_{\odot} \lesssim M_{d,i} \lesssim 3.6~M_{\odot}$ and $0.8~\mathrm{d} \lesssim P_{{\rm orb},i} \lesssim 3~\mathrm{d}$, and most of the candidate systems clustering around $P_{{\rm orb},i} \approx 1 {\rm d}$ with the exception of a few very low $M_{d,i}$. In this figure, the size of the points is set such that a 1 dex increase in the time the model track spent in the allowed parameter space, the size increases by x2. So the size of symbol tracks (non-linearly) the {\it a priori} probability of finding each system in the observed state. However, we make no attempt to correct for population-level effects, such as the initial mass function and the initial mass-ratio and orbital period distributions.

\begin{figure*}
      \centering
      \includegraphics[width=0.8\textwidth]{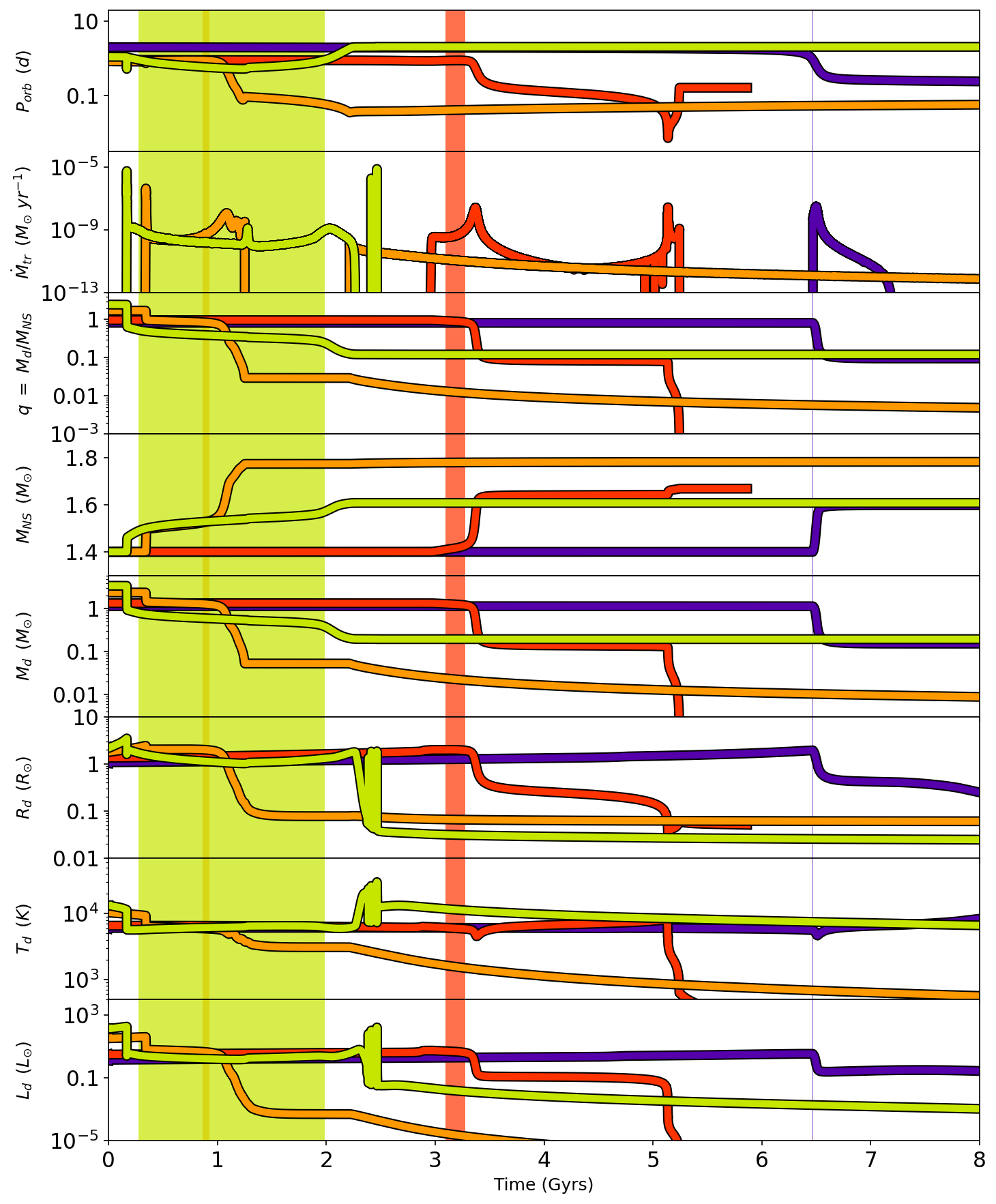}
      \caption[]{Temporal evolution of the relevant donor parameters for the reference evolutionary tracks presented in Figures \ref{fig: param space}--\ref{fig: tracks}. From top to bottom; orbital period ($P_{\rm orb}$), mass transfer rate ($\dot{M}_{tr}$), mass ratio ($q$), Neutron star mass ($M_{\rm NS}$), donor mass ($M_{d}$), radius ($R_d$), temperature ($T_d$), and luminosity ($L_d$), are presented against time.  Each colour correspond to the highlighted model in these figures. The vertical shaded regions span the time each model spends in the allowed region of the parameter space. 
      }
      \label{fig: tracks vs time}
    \end{figure*}

The entire range of allowed initial and present-day donor masses for \src\ is shown in Figure~\ref{fig: Mobs vs Mi}. The present-day donor mass is fairly constrained to lie in the range $0.5~M_{\odot} \lesssim M_{d,i} \lesssim 1.3~M_{\odot}$. The orbital period places \src\ near the so-called "bifurcation period" that separates binaries evolving towards either longer or shorter orbital periods \citep{PylyserSavonije1988AA...191...57P}. This is clear from Figure~\ref{fig: tracks} and \ref{fig: tracks vs time}, where the evolutionary tracks of the compatible sequences are shown. Four distinct "families" of tracks can be identified in this figure. Depending on their initial orbital period an donor mass, systems with the most massive donors will evolve towards longer orbital periods, ending their lives as detached systems (yellow tracks). On the other hand the least massive donor with longer orbital periods will expend most of their lives as detached systems (purple tracks). Systems in between these two (red and orange tracks), follow a converging track. 

 
Figure~\ref{fig: tracks vs time} illustrates the detailed evolution of the system parameters for one example of each type of solution. The yellow tracks show the evolutionary path followed by a diverging system, these are characterised for having initial donor mass of $M_{d,i}\gtrsim 3M_\odot$. In this track, as the donor leaves the main sequence it expands due to nuclear evolution. Once its radius catches up to the Roche Lobe, mass-transfer a relatively long ($\approx 2{\rm Gyr}$) accretion episode is initiated. The system begins this phase as an intermediate mass X-ray binary (IMXB) undergoing rapid, thermal-timescale mass transfer. However, once the mass ratio has been reversed (from $q = M_d/M_{NS} > 1$ to $q < 1$), mass transfer becomes thermally stable and takes place at nearly constant orbital period and donor radius. 
Roughly half-way through this phase, the donor once again approaches the end of its main-sequence lifetime (even at its now lower mass). The orbital period and donor radius then increase again, until the donor loses its envelope in several rapid bursts of mass transfer. The system ultimately ends up as a detached $P_{orb} \simeq 2$~d binary with a NS primary and a low-mass white dwarf secondary.
 
By contrast, the orange tracks show a converging evolution track for a system with $2\lesssim M_{d,i}\lesssim 2.5M_\odot$ donors). Both the initial parameters and early evolution of this system are similar to the diverging track. Here again, mass-transfer begins with an initial TTMT phase after the donor catches up to its Roche Lobe due to nuclear evolution. However, as a result of the very slightly different initial conditions (lower donor mass and shorter orbital period), the donor does not approach the terminal main sequence during the ensuing longer-lived mass-transfer phase. Instead, this phase continues for roughly 1~Gyr, during which donor mass, donor radius and orbital period continually decrease. According to the CARB MB prescription implemented in DABS, MB only effectively stops once the donor mass falls below the Hydrogen-burning limit. The donor then relaxes to its thermal equilibrium radius, and the system detaches. Beyond this point, evolution is driven by GR, which ultimately brings the Roche-lobe back into contact with the donor after about 1~Gyr. The system then continues its slow evolution as an ultra-compact X-ray binary with a sub-stellar companion.

Systems with very low initial donor mass (purple tracks, $M_{d,i}\lesssim 1.5M_\odot$, and long initial orbital periods, $P_{{\rm orb},i}\gtrsim 1{\rm d}$), spend most of their lives as a detached system at long orbital periods. Once the donor leaves the main sequence it expands due to nuclear evolution leading to a relatively short accretion episode. During this phase, the donor loses most of its mass in a short period of time. 

Finally, in systems with low-intermediate donor masses, $M_{d,i}\approx 1.5M_\odot$, but shorter orbital periods (red tracks), mass-transfer is also initiated as a result of donor expansion due to nuclear evolution. However, these systems then go trough a relatively long, stable accretion phase ($\gtrsim 1{\rm Gyr}$). They begin this phase as an intermediate mass X-ray binary (IMXB) undergoing thermal-timescale mass transfer, where most of the donor mass is lost. For the system illustrated in Figure~\ref{fig: tracks vs time}, the stable accretion phase terminates in another brief episode of rapid, unstable mass transfer. During this episode, the donor mass is driven below the sub-stellar limit, and beyond this point mass-transfer The system ultimately reaches very short orbital periods (down to tens of minutes) with a very low (planetary) mass companion. Since such objects {\em expand} in response to adiabatic mass loss, the orbital periods of such systems actually increase again once the thermal time-scale of the donor exceeds the mass-transfer time-scale.

As discussed in Section~\ref{sec:CNO}, the UV line ratios suggest that the donor star is the heavily stripped descendant of a relatively massive donor star \citep[$M_{d,i}\gtrsim 2M_\odot$][]{SchenkerKing2002ASPC..261..242S}. 
This disfavours models with $M_{d,i}\lesssim 2M_\odot$ (i.e. the purple and red tracks). Interestingly, all of the models predict $M_{d,obs} >0.5$, which is in tension with some recent work that favours a low-mass donor in \src\ \citep{Knight2022MNRAS.514.1908K,Vincentelli2023Natur.615...45V}. Instead, the donor properties and evolutionary modelssuggest that the current donor mass is likely to be in the range $0.5\lesssim M_{d,{\rm obs}}\lesssim 1.4M_\odot$. Even with this constraint, \src\ could evolve towards either longer or shorter orbital periods. However, this ambiguity can be further constrained by an accurate determination of the present-day donor mass via a radial velocity study. For instance, a donor mass of $M_{d,{\rm obs}}\gtrsim 1M_\odot$ would point to a shrinking orbit and evolution toward a ultra-compact binary system. If the donor mass is found to be $0.5 M_\odot \lesssim M_{d,{\rm obs}}\lesssim 1M_\odot$, evolution towards a longer orbital period would be favoured (although converging tracks would still be allowed).

\section{Conclusions}

We have presented the far-UV spectrum of the NS-LMXB \src\ across three different accretion states (high-hard/flaring, soft and low-hard states). All of the spectra exhibit anomalous N\,{\sc v}, C\,{\sc iv},  Si\,{\sc iv} and He\,{\sc ii} line ratios, irrespective of the spectral state. This implies that the donor must have been massive enough to produce CNO-processed material. We find that such extreme line ratios can be explained by the equilibrium abundances produced by the CNO cycle for the core temperature of a $\simeq 2 M_\odot$ main-sequence star. This suggests that the initial donor mass fairly massive.

We have also unambiguously detected the donor of \src\ in archival optical observations obtained in quiescence. Modelling the donor's quiescent SED yields $T_{\rm eff,d}\simeq 5700 {\rm K}$ and $R_d\simeq 1.7 R_\odot$, allowing us to place it on the Hertzsprung-Russell diagram. By comparing the observed donor properties to those predicted by theoretical binary evolution calculations, we we are able to determine which evolutionary tracks are consistent with the current state of \src.

We find that all viable evolution tracks start with initial donor masses in the range $1.1~M_{\odot} \lesssim M_{d,i} \lesssim 3.6~M_{\odot}$ and initial orbital periods in the range $0.8~\mathrm{d} \lesssim P_{{\rm orb},i} \lesssim 3~\mathrm{d}$. However, the presence of CNO-processed material suggest a heavily stripped donor with initial mass $M_{d,i} \simeq 2~M_{\odot} - 3.5~M_{\odot}$, which also implies initial orbital periods $P_{{\rm orb},i} \lesssim 1.5~\mathrm{d}$. The initial periods are close to the so-called bifurcation period, so both converging and diverging evolutionary paths are possible for \src.  Diverging tracks would produce a detached system formed with a WD and a NS, while converging track would evolve to an ultra-compact X-ray binary with a sub-stellar companion. All viable evolution tracks predict present-day donor masses in the range $0.5~M_{\odot} \lesssim M_{d,i} \lesssim 1.3~M_{\odot}$. This is in conflict with some recent work favouring a low-mass donor with $M_{d,i} \simeq 0.2~~M_{\odot}$ in \src\ \citep[][]{Knight2022MNRAS.514.1908K,Vincentelli2023Natur.615...45V}. A detailed radial velocity study of the system in quiescence should resolve this tension.

\section*{Acknowledgements}
We thank Boris Gänsicke and Christopher W. Mauche for providing the far-ultraviolet line ratios for cataclysmic variables. We thank Dave Rusell for helpful discussion on the donor nature.
This research made use of Astropy and Matplotlib \citep{Astropy2018AJ....156..123A,matplotlib2007CSE.....9...90H}.

N.C.S. and D.A acknowledges support from the Science and Technology Facilities Council (STFC) grant ST/V001000/1. NCS \& CK also acknowledge support from STFC grant ST/M001326/1.Partial support for KSL's effort on the project was provided by NASA through grant numbers HST-GO-15984 from the Space Telescope Science Institute, which is operated by AURA, Inc., under NASA contract NAS 5-26555. JM acknowledges funding from the Royal Society via a University Research Fellowship.

This work have made use of the Pan-STARRS1 survey. The Pan-STARRS1 Surveys (PS1) and the PS1 public science archive have been made possible through contributions by the Institute for Astronomy, the University of Hawaii, the Pan-STARRS Project Office, the Max-Planck Society and its participating institutes, the Max Planck Institute for Astronomy, Heidelberg and the Max Planck Institute for Extraterrestrial Physics, Garching, The Johns Hopkins University, Durham University, the University of Edinburgh, the Queen's University Belfast, the Harvard-Smithsonian Center for Astrophysics, the Las Cumbres Observatory Global Telescope Network Incorporated, the National Central University of Taiwan, the Space Telescope Science Institute, the National Aeronautics and Space Administration under Grant No. NNX08AR22G issued through the Planetary Science Division of the NASA Science Mission Directorate, the National Science Foundation Grant No. AST-1238877, the University of Maryland, Eotvos Lorand University (ELTE), the Los Alamos National Laboratory, and the Gordon and Betty Moore Foundation.

This research is based on observations made with the NASA/ESA Hubble Space Telescope obtained from the Space Telescope Science Institute, which is operated by the Association of Universities for Research in Astronomy, Inc., under NASA contract NAS 5–26555. These observations are associated with programs 15984 and 16066. Some of the observations reported in this paper were obtained with the Southern African Large Telescope (SALT) under the programme 2018-2-LSP-001 (PI: DAHB). Polish participation in SALT is funded by grant No.\ MEiN nr2021/WK/01. Based on observations obtained with XMM-Newton, an ESA science mission with instruments and contributions directly funded by ESA Member States and NASA.
\section*{Affiliations}
\noindent
{\it
$^{1}$Department of Physics \& Astronomy. University of Southampton, Southampton SO17 1BJ, UK.
\\
$^{2}$Department of Physics, University of Oxford, Denys Wilkinson Building, Keble Road, Oxford, OX1 3RH, UK. \\
$^3$Instituto de Astrof\'isica de Canarias (IAC), E-38205 La Laguna, Tenerife, Spain.				\\
$^4$Departamento de Astrof\'isica, Universidad de La Laguna (ULL), E-38206 La Laguna, Tenerife, Spain.				\\
$^{5}$Space Telescope Science Institute, 3700 San Martin Drive, Baltimore, MD 21218, USA \\
$^{6}$Eureka Scientific, Inc. 2452 Delmer Street, Suite 100, Oakland, CA 94602-3017, USA\\
$^{7}$South African Astronomical Observatory, PO Box 9, Observatory 7935, Cape Town, South Africa.				\\
$^{8}$Southern African Large Telescope, P.O. Box 9, Observatory, 7935, Cape Town, South Africa.\\
$^{9}$Department of Astronomy, University of Cape Town, Private Bag X3, Rondebosch 7701, South Africa.			\\	
$^{10}$ Department of Physics, University of the Free State, PO Box 339, Bloemfontein 9300, South Africa \\
$^{11}$Anton Pannekoek Institute for Astronomy, University of Amsterdam, Science Park 904, 1098 XH, Amsterdam, the Netherlands\\
$^{12}$Independent\\
$^{13}$Center for Astro, Particle and Planetary Physics, New York University Abu Dhabi, PO Box 129188, Abu Dhabi, UAE;  \\
$^{14}$INAF–Osservatorio Astronomico di Brera, Via Bianchi 46, I-23807 Merate (LC), Italy
}
\section*{Data Availability}

The data underlying this article is publicly available in: {\em Swift} \url{https://www.swift.ac.uk/archive/index.php}; {\em Pan-STARRS} \url{https://archive.stsci.edu/panstarrs/}; \url{https://archive.stsci.edu/hst/search.php} program ID 15984 and 16066 for HST data; X-ray data from NICER used all the OBSIDs starting with 120040, 220040,320040 and 359201 accessible from HIESARC (\url{https://heasarc.gsfc.nasa.gov/docs/nicer/nicer_archive.html}); from XMM-Newton Science Archive, OBSID: 0865600201. Remaining data from {\em SALT} will be shared on reasonable request to the corresponding author.



\bibliographystyle{mnras}
\bibliography{My_papers_bio} 




%
%
%
%
%
%

%

\bsp	
\label{lastpage}
\end{document}